\documentclass[pra,11pt,eqsecnum,amsmath,amssymb,superscriptaddress,showpacs,showkeys]{revtex4}

\usepackage{bm,graphicx,hyperref,txfonts}

\newcommand{\rbd}{$^{87\!}{\rm Rb}$}

\newcommand{\crb}{Cram\'{e}r-Rao bound}

\def\fpsi{\hat{\psiup}}

\input{Qcircuit}

\begin{document}

\title{Quantum-limited metrology and Bose-Einstein condensates}

\author{Sergio Boixo}
\affiliation{Institute for Quantum Information, California Institute of Technology,
Pasadena, California 91125, USA}

\author{Animesh Datta}
\affiliation{Institute for Mathematical Sciences, 53 Prince's Gate, Imperial College, London, SW7 2PG, UK}
\affiliation{QOLS, The Blackett Laboratory, Prince Consort Road, Imperial College, London, SW7 2BW, UK}

\author{Matthew J.~Davis}
\affiliation{School of Physical Sciences, University of Queensland, Brisbane,
Queensland 4072, Australia}

\author{Anil~Shaji} \email{shaji@unm.edu}
\affiliation{Department of Physics and Astronomy, University of New Mexico,
Albuquerque, New Mexico 87131-0001, USA}

\author{Alexandre B.~Tacla}
\affiliation{Department of Physics and Astronomy, University of New Mexico,
Albuquerque, New Mexico 87131-0001, USA}

\author{Carlton M.~Caves}
\affiliation{Department of Physics and Astronomy, University of New Mexico,
Albuquerque, New Mexico 87131-0001, USA}

\begin{abstract}
We discuss a quantum-metrology protocol designed to estimate a
physical parameter in a Bose-Einstein condensate of $N$ atoms, and we
show that the measurement uncertainty can decrease faster than $1/N$.
The $1/N$ scaling is usually thought to be the best possible in any
measurement scheme. From the perspective of quantum information
theory, we outline the main idea that leads to a measurement
uncertainty that scales better than $1/N$. We examine in detail some
potential problems and challenges that arise in implementing such a
measurement protocol using a Bose-Einstein condensate. We discuss how
some of these issues can be dealt with by using lower-dimensional
condensates trapped in nonharmonic potentials.
\end{abstract}



% 03.65.Ta  Foundations of quantum mechanics; measurement theory
% 03.65.-w  Quantum mechanics
% 03.75.Nt  Other Bose-Einstein condensation phenomena
% 03.75.Mn  Multicomponent condensates; spinor condensates
\maketitle

\section{Introduction}
\label{sec:intro}

In quantum metrology, the description ``Heisenberg-limited scaling"
refers to the best possible scaling of the measurement uncertainty
with the resources put into a measurement. The phrase arises not from
Heisenberg uncertainty relations, but from uncertainty relations of
the Mandelstam-Tamm type~\cite{mandelstam45a},
\begin{equation}
    \label{eq:mt}
    \delta \gamma \langle \Delta^2 K\rangle^{1/2} \geq \frac{1}{2}\;,
\end{equation}
in units with $\hbar =1$. The uncertainty $\delta \gamma$ in a
parameter $\gamma$ that, in part, determines the state of a quantum
system is related to the standard deviation of the operator $K$ that
generates translations of the state along a path parameterized by
$\gamma$.  A sequence of logical and mathematical steps is needed to
provide a rigorous connection between the problem of measurement
uncertainty in quantum metrology and uncertainty relations of the
Mandelstam-Tamm type. The pioneering work of
Helstrom~\cite{helstrom76a}, Holevo~\cite{holevo82a}, Braunstein,
Caves, and Milburn~\cite{braunstein94a,braunstein96a}, and others
laid out and elucidated these steps. We summarize them below for the
sake of completeness.

The discussion in this paper is restricted to single-parameter
estimation. The first step in estimating the value of a parameter is
to identify an elementary physical system that is sensitive to
changes in the parameter, just as one would choose a balance to
measure weight or a thermometer to measure temperature. One or more of
these elementary systems make up the measuring device or {\em probe}.
The measurement uncertainty is a property of this measuring device.
In quantum metrology this means that we expect the measurement
uncertainty to depend on the initial state of the quantum probe, its
evolution, and the measurement made on the probe to extract
information about the parameter. The {\em quantum \crb\/} quantifies
the idea that the optimal measurement uncertainty is inversely
proportional to the change in the state of the probe corresponding to
small changes in the value of the parameter:
\begin{equation}
    \label{eq:qcrb} (\delta \gamma)^2 \geq \frac{1}{(ds_{\rm DO}/d \gamma)^2} =
    \frac{1}{{\mathfrak{I}}(\gamma, t)}\;.
\end{equation}
Here $ds_{\rm DO}$ denotes a distance element in the space of density
operators of the probe, and ${\mathfrak{I}}(\gamma, t)$ is the
quantum Fisher information. The uncertainty in determining $\gamma$
is quantified by the units-corrected, root-mean-square deviation of
one's estimate of the parameter, $\gamma_{\rm est}$, from the true
value~$\gamma$:
\begin{equation}
    \label{eq:deltagamma}
    \delta \gamma = \bigg \langle \bigg( \frac{\gamma_{\rm est}} {
    |\partial\langle \gamma_{\rm est} \rangle /\partial \gamma |} -
    \gamma \bigg)^2 \bigg\rangle^{\!1/2}\;.
\end{equation}

In classical statistics, the \crb\ on measurement uncertainty is
given by
\begin{equation}
    \label{eq:classicalCR}
    (\delta\gamma)^2\ge\frac{1}{\mathcal{I}(\gamma)}\;,
\end{equation}
where
\begin{equation}
    \label{eq:fisher1} {\mathcal I}(\gamma) \equiv
    \Bigg\langle \bigg(
    \frac{\partial \;}{\partial \gamma} \ln p(\zeta|\gamma) \bigg)^{\!2} \Bigg \rangle\;,
\end{equation}
called the {\em Fisher information}, is an average over the
probability distribution $p(\zeta|\gamma)$ for a random variable
$\zeta$ and is a measure of the information that $\zeta$ can provide
about $\gamma$.  The classical \crb~(\ref{eq:classicalCR}) can
generally be achieved only asymptotically in a large number of
trials, i.e., independent measurements of $\zeta$.  The requirement
of many trials to achieve the \crb\ is important, but as a purely
classical effect, it is not germane to our discussion of quantum
limits, so we do not consider it further in the remainder of this
paper.

If the probe used to estimate the value of $\gamma$ were a classical
system, then $\zeta$ would label the possible states of the probe at
the end of the measurement, with $p(\zeta|\gamma)$ being the
probability of finding the probe in each of these states. For a
quantum probe in a state $\rho(\gamma,t)$ at the end of the
measurement process, $\zeta$ labels the possible outcomes of a
measurement performed on the probe, which is described by POVM
elements $E(\zeta)$, with $\int d \zeta \,
E(\zeta)=\openone$ and $p(\zeta|\gamma, t)
={\mbox{tr}}[E(\zeta) \rho(\gamma, t)]$.  The classical Fisher
information ${\mathcal I}(\gamma,t)$, defined using
$p(\zeta|\gamma,t)$, clearly depends on the choice of POVM.  The
quantum Fisher information, which is independent of the choice of
POVM, is therefore defined as
\begin{equation}
    \label{eq:fisher2}
    {\mathfrak{I}}(\gamma,t) \equiv
    \max_{E(\zeta)} \,{\mathcal I}(\gamma,t)\;.
\end{equation}
The maximization over all possible measurements in the above equation
is a rather daunting prospect, but it can be shown
that~\cite{helstrom76a,holevo82a,braunstein94a,braunstein96a}
\begin{equation}
    \label{eq:fisher3} {\mathfrak{I}}(\gamma,t) =
    {\mbox{tr}} \big[ \rho(\gamma,t) {\mathfrak{L}}^2(\gamma,t)\big] =
    \big\langle \mathfrak{L}^2(\gamma, t) \big\rangle\;.
\end{equation}
The symmetric logarithmic derivative, $\mathfrak{L}(\gamma, t)$, is
the Hermitian operator defined implicitly by the equation
\begin{equation}
    \label{eq:symlog}
    \frac{1}{2}\big( \mathfrak{L}\rho + \rho \mathfrak{L} \big) =
    \frac{\partial \rho}{\partial \gamma}\;.
\end{equation}

We now make two simplifying assumptions.  First, we assume that
translations in the parameter are generated by a unitary operator.
This allows us to characterize the translations in terms of a
Hermitian generator $K(\gamma, t)$ defined by
\begin{equation}
    \label{eq:Kgen}
    \frac{\partial \rho(\gamma,t)}{\partial \gamma}
    = -i \big[K(\gamma, t), \, \rho(\gamma, t) \big]\;.
\end{equation}
Second, we assume that the state of the probe is pure, which implies
$\rho^2(\gamma,t)=\rho(\gamma,t)$. Under these two assumptions, we
can identify the symmetric logarithmic derivative as
\begin{equation}
    \mathfrak{L}(\gamma,t) =
    2\frac{\partial\rho}{\partial\gamma} =
    -2i\big[K(\gamma,t), \, \rho(\gamma,t)\big]\;,
\end{equation}
and the quantum Fisher information becomes
\begin{equation}
    \mathfrak{I}(\gamma,t) =
    4\,{\mbox{tr}}\big(\rho K^2 - \rho K\rho K \big) =
    4 \langle \Delta^2 K(\gamma,t) \rangle\;.
\end{equation}
Thus, using Eq.~(\ref{eq:qcrb}), we obtain
\begin{equation}
    \label{eq:fisher3a}
    \delta \gamma \geq \frac{1}{2 \langle \Delta^2 K(\gamma,t) \rangle^{1/2}}\;,
\end{equation}
which is a rigorous statement of the Mandelstam-Tamm uncertainty
relation~(\ref{eq:mt}). Our two simplifying assumptions can be
relaxed~\cite{helstrom76a,holevo82a,braunstein94a,braunstein96a}, but
we do not need the more general forms of the \crb\ in this paper.

We can further simplify Eq.~(\ref{eq:fisher3a}) by noting that the
variance of a Hermitian operator is bounded from above by $\langle
\Delta^2 K \rangle \leq \|K\|^2/4$, where $\| \, \cdot \, \|$ is
defined as the difference between the largest and smallest
eigenvalues of a Hermitian operator (this is a semi-norm for
Hermitian operators).  The quantum \crb\ then becomes
\begin{equation}
    \label{eq:qcrb2} \delta \gamma \geq \frac{1}{\|K(\gamma, t)\|}\;.
\end{equation}

We mentioned the elementary quantum systems, sensitive to $\gamma$,
that are used to build the probe. The number $N$ of such elementary
units of the probe can be regarded as the most significant resource
that goes into a measurement scheme. The differences between the
tensor-product state space of a composite quantum system of $N$ probe
units and the Cartesian-product state space of an equivalent,
classical composite system is the motivation for investigating
whether a composite quantum probe offers advantages over classical
ones in the relationship between $\delta \gamma$ and $N$.

From Eq.~(\ref{eq:qcrb2}) we see that theoretically the
$N$-dependence of the bound on $\delta \gamma$ comes solely from the
dependence of the generator $K$ on $N$. In any particular
quantum-metrology scheme, however, this bound might not be
achievable, and additional dependence of $\delta \gamma$ on $N$ can
come from the nature of the state of the probe as well. To see these
dependences clearly and to understand what ``Heisenberg-limited
scaling" means, we view quantum metrology from the perspective of
quantum information theory using the language of quantum circuits in
Sec.~\ref{sec:metrology}. We also explain how one can construct
measurement protocols in which $\delta \gamma$ scales with $N$ in a
manner not thought to be possible until recently.
Section~\ref{sec:bec} examines in some detail how such an enhanced
metrology protocol might be implemented in a Bose-Einstein condensate
(BEC) of $N$ atoms and considers the various problems and issues that
might arise in a BEC realization of the proposed metrology scheme.

\section{Quantum metrology from an information-theoretic perspective
\label{sec:metrology}}

In this section, we follow Giovannetti, Lloyd, and
Maccone~\cite{giovannetti06a} in using quantum circuits to describe
and analyze metrology protocols.  From this perspective, we first
look at a couple of well-known measurement schemes that were
considered in~\cite{giovannetti06a}---Ramsey interferometry
(Sec.~\ref{subsec:ramsey}) and interferometry using a
Schr\"odinger-cat state (superposition of macroscopically distinct
states) (Sec.~\ref{subsec:cat})---with the aim of generalizing these
circuits to new protocols that were introduced
in~\cite{boixo07a,boixo08a}. In these initial discussions of Ramsey
interferometry and cat-state interferometry, we assume that the
elementary quantum systems that make up the probe are qubits. The
quantization axis is taken to be along the $z$-direction of a
Bloch-sphere representation, with the standard basis states along
this direction denoted as $|0\rangle$ and $|1 \rangle$. Despite the
notation the actual qubits need not be spin-$1/2$ particles; they
could very well be atoms in which only two energy levels are relevant
or a variety of other suitable systems. In the subsequent general
discussions of linear and nonlinear interferometry
(Secs.~\ref{subsec:linear} and \ref{subsec:nonlinear}), we allow the
probe units to be any quantum system.  It turns out, however, that
optimal sensitivities are always attained by using only two levels of
each unit, so in the end we can always regard the probe units as
qubits.

In all the quantum circuits depicted in this section, we use $N=3$
probe units as an example.

\subsection{Ramsey interferometry \label{subsec:ramsey}}

\begin{figure}
\begin{equation*}
    \Qcircuit @C=1.4em @R=1.3em {
    \lstick{\ket{0}} & \qw & \gate{H} & \gate{U_{\varphi}=e^{-i \sigma_z \varphi/2}} & \gate{H} & \measure{M_z} & \cw \\
    \lstick{\ket{0}}& \qw & \gate{H} & \gate{U_{\varphi}=e^{-i \sigma_z \varphi/2}} & \gate{H} & \measure{M_z} & \cw \\
    \lstick{\ket{0}}& \qw & \gate{H} & \gate{U_{\varphi}=e^{-i \sigma_z \varphi/2}} & \gate{H} & \measure{M_z} & \cw }
\end{equation*}
\caption{Quantum circuit for Ramsey interferometry.\label{fig:Ramsey}}
\end{figure}
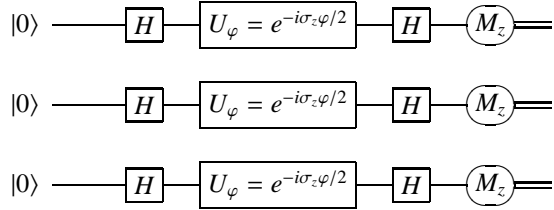

A typical Ramsey interferometer, such as the one
in~\cite{gleyzes07a}, can be represented by the quantum circuit in
Fig.~\ref{fig:Ramsey}.  In this measurement protocol, each of the $N$
qubits that make up the probe evolves independently.  All the qubits
are initialized in the state $|0\rangle$, which might represent the
ground state in Ramsey interferometry using atoms.  The Hadamard gate
$H$ puts each of the qubits in an equal superposition of the two
basis states, $(|0\rangle + |1\rangle)/\sqrt{2}$. The
parameter-dependent evolution of the quantum probe is generated by
the Hamiltonian
\begin{equation}
    \label{eq:ramseyH} H_{\rm Ramsey} =
    \gamma \sum_{j=1}^N \sigma_{z;\,j}/2 = \gamma J_{z}\;,
\end{equation}
where $\sigma_{z;\,j}$ denotes the $\sigma_z$ operator acting on the
$j$th probe qubit and $J_{z}$ is the $z$ component of the ``total
angular momentum'' for all the qubits.

Evolution under this Hamiltonian for a time $t$ introduces a relative
phase $\varphi \equiv \gamma t$ between the two components of the
superposition, changing the state of the probe qubits to
$(e^{-i\varphi/2}|0\rangle + e^{i \varphi/2}|1 \rangle)/\sqrt{2}$.
The last set of Hadamard gates changes the parameter-dependent phases
in the superpostion into amplitude (population) information.  Thus
the state of the probe qubits just before the readout is
$\cos(\varphi/2)|0\rangle + \sin(\varphi/2)|1 \rangle$. The final
readout in Ramsey interferometry is done by measuring each of the
qubits along the $z$-direction.  This leads to a measured signal
\begin{equation}
    \label{eq:ramseySIG}
    \langle J_{z} \rangle \equiv
    \Biggl\langle \frac{1}{2} \sum_{j=1}^N \sigma_{z;\,j} \Biggr\rangle =
    \frac{1}{2} N \cos \varphi\;.
\end{equation}
The variance in the signal is
\begin{equation}
    \label{eq:ramseyVAR}
    \langle \Delta^2 J_{z} \rangle =
    \frac{1}{4} N \langle \Delta^2 \sigma_z \rangle =
    \frac{1}{4} N \sin^2\!\varphi\;.
\end{equation}
The uncertainty in the estimate of $\gamma$ from the measured signal
in~Eq.~(\ref{eq:ramseySIG}) is
\begin{equation}
    \label{eq:ramseyUN}
    \delta \gamma_{\rm Ramsey} =
    \frac{\langle \Delta^2 J_{z} \rangle^{1/2}}
    {\big|d \langle J_{z} \rangle/d\gamma\big|} = \frac{1}{t\sqrt{N}}\;.
\end{equation}

For the Ramsey Hamiltonian~(\ref{eq:ramseyH}), the generator of
translations in $\gamma$ is $t J_{z}$.  Thus, according to the quantum
\crb~(\ref{eq:qcrb2}), the measurement uncertainty is bounded from
below by
\begin{equation}
    \delta \gamma \geq \frac{1}{t\|J_z\| } =
    \frac{1}{tN\|\sigma_z/2\|} = \frac{1}{tN}\;.
\end{equation}
The Ramsey interferometer described here does not achieve the best
measurement uncertainty given by the quantum \crb. The Hamiltonian
$H_{\rm Ramsey}$ that governs the evolution of the probe qubits is
fixed by the choice of physical systems that are the qubits. Given a
choice of probe qubits, however, we still have the freedom to choose
an optimal initial state for the probe and an optimal measurement of
the qubits to minimize the measurement uncertainty. It turns out that
the best possible scaling for the measurement uncertainty can be
achieved if the probe is initialized in an entangled,
``Schr\"{o}dinger-cat" state~\cite{bollinger96a,huelga97a}.

\subsection{Cat-state interferometry \label{subsec:cat}}

\begin{figure}
\begin{equation*}
    \Qcircuit @C=1.4em @R=1.3em {
    \lstick{\ket{0}}& \gate{H}  &  \ctrl{1} & \ctrl{2}  &  \gate{U_{\varphi}}  & \ctrl{2}  & \ctrl{1}  & \gate{H}  & \measure{M_z} & \cw \\
    \lstick{\ket{0}}& \qw       & \targ     & \qw       &  \gate{U_{\varphi}}  & \qw       & \targ     &\qw        & \qw           & \qw \\
    \lstick{\ket{0}}& \qw       & \qw       & \targ     &  \gate{U_{\varphi}}  & \targ     & \qw       & \qw       & \qw           & \qw
    }
\end{equation*}
\caption{Quantum circuit for cat-state interferometry.\label{fig:cat}}
\end{figure}
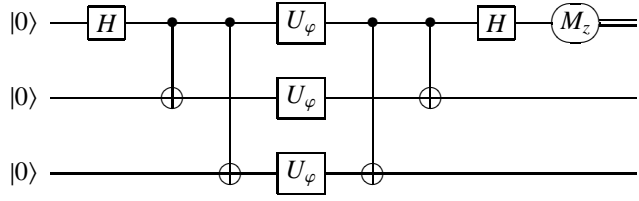

The quantum circuit that uses a probe initialized in a
Schr\"{o}dinger-cat state is depicted in Fig.~\ref{fig:cat}. The
Hadamard gate on the first qubit, followed by the controlled-NOT
gates to the remaining qubits, initializes the probe in the state
$|\mbox{cat}\rangle = \big(|0\ldots 0\rangle + |1\ldots 1\rangle
\big)/\sqrt2$. This state is often referred to as the
Schr\"{o}dinger-cat state because when the number of qubits is large,
it is a superposition of two macroscopically distinct states.

The probe qubits evolve under the same Hamiltonian~(\ref{eq:ramseyH})
as in Ramsey interferometry. The parameter-dependent evolution of the
probe for a duration $t$ changes the probe state to $\big(
e^{-iN\varphi/2}|0\ldots 0\rangle + e^{iN\varphi /2}| 1 \ldots
1\rangle \big)/\sqrt2$, where $\varphi = \gamma t$. After the
parameter-dependent evolution, one way to handle the readout,
depicted in the circuit above, is to subject the qubits to a sequence
of gates that kick the phases picked up by the two components of the
cat state into amplitudes on the first qubit, so that the state of
the probe just before readout is $[\cos (N \varphi /2)\, |0 \rangle +
\sin (N \varphi /2) \,| 1\rangle] \otimes | 0
\rangle^{\otimes(N-1)}$. The readout of the probe can then be
performed by measuring the $\sigma_z$ operator on the first qubit.
This leads to a measured signal and variance given by
\begin{equation}
    \label{eq:catsigvar}
    \langle \sigma_{z;1} \rangle = \cos N \varphi \quad\mbox{and}\quad
    \langle \Delta^{2} \sigma_{z;1} \rangle  =\sin^2 N\varphi \;.
\end{equation}
The frequency of the $\gamma$-dependent fringe in cat state
interferometry is $N$ times greater than the frequency of the signal
in ordinary Ramsey interferometry. This leads to an enhanced
sensitivity in the estimate of $\gamma$ in cat-state interferometry,
which achieves the \crb:
\begin{equation}
    \label{eq:catun}
    \delta \gamma_{\rm cat} =
    \frac{\langle \Delta^2 \sigma_{z;1} \rangle^{1/2}}
    {\big|d \langle \sigma_{z;1} \rangle/d\gamma\big|} =
    \frac{1}{t N}\;.
\end{equation}

\subsection{Heisenberg-limited metrology with linear Hamiltonians \label{subsec:linear}}

We can put our interferometry circuits in a general setting by
considering the case in which the probe units are arbitrary systems
and the probe Hamiltonian is of the form
\begin{equation}
    \label{eq:linearH}
    H_{\rm linear} = \gamma h_{\rm linear} = \gamma \sum_{j=1}^{N} h_{j}\;.
\end{equation}
Here the operators $h_j$ denote identical couplings to the probe
units; the use of independent couplings to the parameter is the
source of our appelation ``linear'' for this Hamiltonian.  The
generator of translations in $\gamma$ is $K(\gamma,t)=th_{\rm
linear}$, so the quantum \crb~(\ref{eq:qcrb2}) on the uncertainty in
a determination of $\gamma$ takes the form
\begin{equation}
    \label{eq:linearCRB}
    \delta \gamma \geq \frac{1}{t\|h_{\rm linear} \|} =
    \frac{1}{tN(\Lambda - \lambda)}\;,
\end{equation}
where $\Lambda$ and $\lambda$ are the largest and smallest
eigenvalues of the single-unit operators $h_j$.  Achieving the \crb\
only requires using two levels of each unit, the eigenstates
$|\Lambda\rangle$ and $|\lambda\rangle$ corresponding to the largest
and smallest eigenvalues of the operators $h_j$, so we can always
regard the units as qubits with $|0\rangle=|\Lambda\rangle$ and
$|1\rangle=|\lambda\rangle$.

\begin{figure}
\begin{equation*}
    \Qcircuit @C=1.4em @R=1.3em {
                            &       & \lstick{\ket{S}}  & \multigate{2}{P}  &  \gate{U_{\varphi}=e^{-ih_1\varphi}}  &  \multigate{3}{R} & \measure{M}   & \cw \\
                            &       & \lstick{\ket{S}}  & \ghost{P}         &  \gate{U_{\varphi}=e^{-ih_2\varphi}}  &  \ghost{R}        & \measure{M}   & \cw \\
                            &       & \lstick{\ket{S}}  & \ghost{P}         &  \gate{U_{\varphi}=e^{-ih_3\varphi}}  &  \ghost{R}        & \measure{M}   & \cw \\
     \lstick{{\rm ancilla}} & \qw   &\qw                & \qw               &  \qw                                  &  \ghost{R}        & \measure{M}   & \cw
     \gategroup{1}{1}{3}{4}{.8em}{--}
     \gategroup{1}{5}{3}{5}{.8em}{--}
     \gategroup{1}{6}{4}{8}{.8em}{--}
    }
\end{equation*}
\caption{Quantum circuit for a general linear
interferometer.\label{fig:linearint}}
\end{figure}
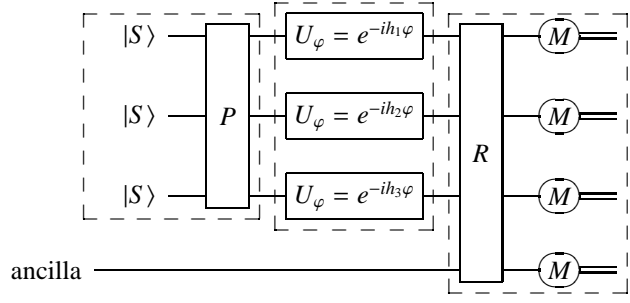

The quantum circuit that represents a measurement protocol of this
sort is drawn in Fig.~\ref{fig:linearint}.  The dashed boxes
highlight the three stages of this protocol: probe preparation,
dynamics, and readout.  All the probe units begin in a standard state
$|S\rangle$. The arbitrary unitary operator $P$ can then prepare any
initial state as input to the dynamics. In the dynamics stage, the
gates $U_\varphi$ imprint information about the parameter on the
probe. The final readout stage includes an arbitrary unitary
interaction $R$ among the probe units and with an arbitrary ancilla
system.  This unitary followed by measurements on each subsystem in a
standard basis can be used to perform any quantum measurement.  The
quantum \crb~(\ref{eq:linearCRB}) applies to all circuits of the
above form. Indeed, the bound actually applies to somewhat more
general situations in which the unitary operator $R$ is interleaved
with the gate dynamics and the results of ancilla measurements are
fed back onto the probe~\cite{boixo07a}.

If the preparation unitary $P$ is omitted from the circuit, making
the input to the dynamics a product state, then the uncertainty in
the generator of $\gamma$ displacements is bounded by
$\langle\Delta^2 K\rangle^{1/2}\le t\sqrt N(\Lambda-\lambda)/2$.  The
resulting bound on measurement uncertainty, from
Eq.~(\ref{eq:fisher3a}), is
\begin{equation}
\label{eq:qnl}
\delta\gamma\ge\frac{1}{t\sqrt N(\Lambda-\lambda)}\equiv\delta\gamma_{\rm QNL}\;.
\end{equation}
This bound, a general form of that for standard Ramsey
interferometry, is called the {\em quantum noise limit\/} (QNL) or
the {\em shot-noise limit}. The optimal $1/t\sqrt N$ sensitivity for
product-state inputs can be achieved by using initial state
$|S\rangle=\big(|\Lambda\rangle+|\lambda\rangle\big)/\sqrt2$ for each
unit and by making a final product measurement of an equatorial-plane
spin component on each unit (in the qubit Bloch sphere formed from
$|0\rangle=|\Lambda\rangle$ and $|1\rangle=|\lambda\rangle$).

One can achieve the \crb~(\ref{eq:linearCRB}) by operating the
circuit in a way that takes advantage of entangled input states.  The
preparation operator is chosen to take the initial product of
standard states to the ``cat-like'' state $\big( |\Lambda, \ldots,
\Lambda \rangle + |\lambda, \ldots, \lambda \rangle \big)/\sqrt2$. In
the dynamics stage, this ``cat-like'' initial state is subject to a
period of parameter-dependent evolution that changes it to $\big(
e^{-iN\Lambda \varphi}|\Lambda, \ldots, \Lambda \rangle +
e^{iN\lambda \varphi}|\lambda, \ldots, \lambda \rangle \big)/\sqrt2$.
The readout process kicks back the differential phase shift into
amplitude information, which produces fringes with frequency
proportional to $N(\Lambda - \lambda)$, thus achieving the optimal
measurement uncertainty,
\begin{equation}
\delta \gamma=\frac{1}{tN(\Lambda - \lambda)}
\equiv\delta\gamma_{\rm HL}\;,
\end{equation}
of the \crb~(\ref{eq:linearCRB}).  This optimal measurement
uncertainty, a general form of that for cat-state interferometry, is
often called the {\em Heisenberg limit}.

The general quantum-metrology scheme considered in this subsection
indicates that probe preparation gives an enhancement of $1/\sqrt{N}$
over the case where the probe qubits are initialized in a product
state.  Readout has already been optimized to take advantage of this
entangled input, so we conclude that when the parameter-dependent
dynamics acts independently on the probe qubits, Heisenberg-limited
scaling is indeed the $1/N$ scaling.  {\em The one remaining way of
exploring whether the $1/N$ scaling can be improved is to consider
more general
dynamics}~\cite{luis04a,beltran05a,boixo07a,luis07a,rey07a,boixo08a,choi07a,woolley08a};
we turn to that possibility in the next subsection.

\subsection{Heisenberg-limited metrology with nonlinear Hamiltonians \label{subsec:nonlinear}}

A generalized quantum-metrology scheme in which the dynamics of the
probe is generated by a Hamiltonian that includes all $k$-body
couplings between the probe qubits was first considered
in~\cite{boixo07a}.  This nonlinear coupling Hamiltonian has the form
\begin{equation}
    \label{eq:nonlinH}
    H_{\rm nonlinear} = \gamma h_{\rm nonlinear}
    = \gamma\,\biggl(\sum_{j=1}^N h_j\biggr)^k =
    \gamma \sum_{j_{1}, \ldots, j_{k}=1}^{N} h_{j_{1}}h_{j_{2}}\cdots h_{j_{k}}\;.
\end{equation}
The generator of translations in $\gamma$ is $K(\gamma,t)=th_{\rm
nonlinear}$, so the quantum \crb\ for this dynamics is
\begin{equation}
    \label{eq:nonlincrb}
    \delta \gamma \geq \frac{1}{tN^{k}(\Lambda_{\rm max}-\Lambda_{\rm min})}\;,
\end{equation}
where $\Lambda_{\rm max}$ and $\Lambda_{\rm min}$ are functions of
$\Lambda$ and $\lambda$, the largest and smallest eigenvalues,
respectively, of the single-unit operators $h_j$. For instance, if
both $\Lambda$ and $\lambda$ are positive, then $\Lambda_{\rm max} =
\Lambda^{k}$ and $\Lambda_{\rm min} = \lambda^{k}$ for all values of
$k$.  The other possible signs of $\Lambda$ and $\lambda$ are
discussed in~\cite{boixo08a}; they all lead to a $1/N^k$ scaling.

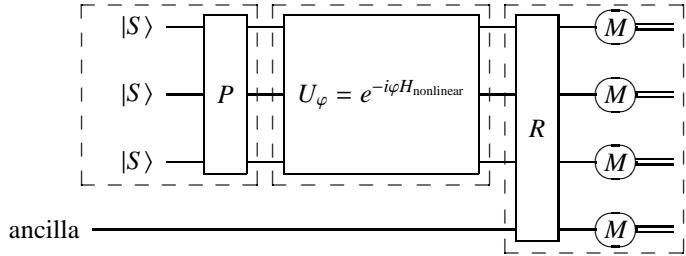
\begin{figure}
\begin{equation*}
    \Qcircuit @C=1.4em @R=1.3em {
                            &       & \lstick{\ket{S}}  & \multigate{2}{P}  &  \multigate{2}{U_\varphi=e^{-i\varphi H_{\rm nonlinear}}} &  \multigate{3}{R} & \measure{M}   & \cw \\
                            &       & \lstick{\ket{S}}  & \ghost{P}         &  \ghost{U_\varphi=e^{-i\varphi H_{\rm nonlinear}}}        &  \ghost{R}        & \measure{M}   & \cw \\
                            &       & \lstick{\ket{S}}  & \ghost{P}         &  \ghost{U_\varphi=e^{-i\varphi H_{\rm nonlinear}}}        &  \ghost{R}        & \measure{M}   & \cw \\
     \lstick{{\rm ancilla}} & \qw   &\qw                & \qw               &  \qw                                                      &  \ghost{R}        & \measure{M}   & \cw
     \gategroup{1}{1}{3}{4}{.8em}{--}
     \gategroup{1}{5}{3}{5}{.8em}{--}
     \gategroup{1}{6}{4}{8}{.8em}{--}
    }
\end{equation*}
\caption{Quantum circuit for a general nonlinear
interferometer.\label{fig:nonlinearint}}
\end{figure}

The quantum circuit for metrology with nonlinear Hamiltonians has the
form shown in Fig.~\ref{fig:nonlinearint}.  This circuit has the same
overall form as that for linear quantum metrology, with the same
three stages highlighted by the dashed boxes.  The only difference
comes in the dynamics stage, where the gate that imprints information
about the parameter on the probe involves simultaneous coupling to
all the probe units.

To achieve the $1/N^{k}$ scaling made available by using a nonlinear
Hamiltonian, the probe units have to be initialized in an entangled
state that is very much like a cat state.  Experimental limitations
up till now have precluded making such cat-like states for large
numbers of systems.  To avoid this difficulty, Boixo~{\em et
al.}~\cite{boixo08a} analyzed the performance of quantum-metrology
protocols employing nonlinear Hamiltonians when the initial state of
the probe is a product state.  In this case the optimal measurement
uncertainty scales as
\begin{equation}
    \label{eq:productCRB}
    \delta \gamma \sim \frac{1}{tN^{k-1/2}}\;.
\end{equation}
The factors multiplying this scaling depend on the particular
nonlinear coupling Hamiltonian~\cite{boixo08a}.  It is noteworthy
that the optimal $1/tN^{k-1/2}$ sensitivity can be achieved using
product measurements of equatorial spin components in the effective
qubit space formed from $|\Lambda\rangle$ and $|\lambda\rangle$.  The
key point is that for a $k$-body coupling Hamiltonian, the use of a
product-state input costs only a factor of $\sim\!\!\sqrt{N}$ relative to
the optimal sensitivity~(\ref{eq:nonlincrb}).  The quantum noise
limit and the Heisenberg limit of linear metrology are a special case
of this $\sqrt N$ loss of sensitivity when using input product states
as opposed to an optimal entangled state.

With two-body couplings and an initial product state for the probe, a
measurement uncertainty scaling as $1/N^{3/2}$ is possible.  Since
two-body coupling between all probe units is not an especially
onerous requirement for a probe system, the prospect of improving
upon the $1/N$ Heisenberg scaling motivates us to investigate
candidate systems for such metrology schemes.  In the next section we
consider a Bose-Einstein condensate (BEC) as such a candidate system
with the aim of developing a detailed, realistic, and viable proposal
for an experiment that achieves better than $1/N$ scaling for the
measurement uncertainty in quantum single-parameter estimation.

\subsection{Role of entanglement}
\label{subsec:entanglement}

The more general point of view provided by nonlinear quantum
metrology allows us to see exactly what benefit entanglement bestows
on quantum metrology.  Entanglement permits one to marshal the
available resources in a quantum-metrology protocol into an initial
state that can achieve the best possible scaling for the measurement
uncertainty as laid out by the quantum \crb.  The use of an
appropriately entangled input state purchases a sensitivity boost by
a factor of $\sim\!1/\sqrt{N}$ relative to the use of an optimal
initial product state.  Initial entanglement is, however, not
necessary for getting to or improving upon the $1/N$ Heisenberg
scaling.

When specialized to qubits, with $h_j=\sigma_{z;\,j}/2$, and to
quadratic couplings, the nonlinear probe Hamiltonian of
Eq.~(\ref{eq:nonlinH}) becomes $H_{\rm nonlinear}=\gamma
J_z^{\,2}$. This Hamiltonian generates entanglement during the
dynamics stage of the protocol.  Despite the evidence from the
quantum \crb\ that entanglement only helps in the initial state, one
might reasonably ask whether this dynamically generated entanglement
plays a role in improving upon the $1/N$ scaling.  The
$J_z^{\,2}$ probe Hamiltonian was analyzed in detail
in~\cite{boixo08a}.  The optimal initial product state is $[\cos
(\pi/8)|0\rangle + \sin(\pi/8)|1\rangle]^{\otimes n}$. Evolution
under the $J_z^{\,2}$ Hamiltonian for a short time
$t\ll\gamma^{-1}$, followed by a measurement of $J_y$, gives a
measurement precision $\delta\gamma =2/N^{3/2}$, which is the optimal
precision for this Hamiltonian and initial state.  As a consequence
of the $J_z^{\,2}$ evolution, however, the probe qubits become
entangled and suffer from an associated ``phase dispersion'' that
makes the measurement uncertainty large for separable measurements
when $\gamma t$ becomes large.  Far from being an aid, the generated
entanglement seems only to make it impossible to achieve the
$1/N^{3/2}$ sensitivity using product measurements.

In~\cite{boixo08b} it was pointed out that in addition to the
$J_{z}^{\,2}$ Hamiltonian, $H_{\rm nonlinear} = \gamma N
J_{z}$, can also be used to get an optimal measurement
uncertainty $1/N^{3/2}$ when the probe qubits all start off in the
state $(|0\rangle+|1\rangle)/\sqrt2$.  The $N J_{z}$ Hamiltonian
does not produce entanglement between the probe qubits, nor does it
produce phase dispersion.  Indeed, the $N J_{z}$ Hamiltonian
acts like a linear coupling whose strength is enhanced by a factor of
$N$. In this case it is clearly the {\em dynamics\/} alone that leads
to the enhanced scaling for the measurement uncertainty, since there
is no entanglement between the probe qubits at any stage in the
metrology protocol.

On physical grounds, the $N J_z$ Hamiltonian cannot be a fundamentally
linear coupling whose strength is enhanced by addition of qubits, but
rather must come naturally from quadratic couplings to the parameter.
Such a coupling does appear in the Hamiltonian for a two-mode BEC, as
was pointed out in~\cite{boixo08b}.  We introduce this BEC
implementation in the next section and discuss it in some detail.

\section{Bose-Einstein condensate as a quantum probe \label{sec:bec}}

\subsection{Nonlinear BEC interferometry}

The many-body Hamiltonian for a dilute Bose gas consisting of atoms
of mass $m$ in a trapping potential $V(\bm{r})$ at zero temperature,
in second-quantized notation, is given
by~\cite{leggett01a,fetter99a,dalfovo99a,rogelsalazar04a}
\begin{equation}
\label{BEChamil1}
\hat H =
\int d{\bm r}\,
\left( \frac{\hbar^2}{2 m } \nabla \fpsi^{\dagger}  \cdot \nabla \fpsi+
V({\bm r})\fpsi^{\dagger} \fpsi +
\frac{1}{2}g\fpsi^{\dagger} \fpsi^{\dagger} \fpsi \fpsi \right)\;,
\end{equation}
where $\fpsi^{\dagger}({\bm r})$ and $\fpsi({\bm r})$ are creation
and annihilation field operators that obey bosonic commutation
relations,
\begin{equation}
\big[ \fpsi({\bm r}), \, \fpsi^{\dagger} ({\bm r}')\big]  =
\delta^{(3)} ({\bm r} - {\bm r}'), \quad \big[ \fpsi({\bm r}), \,
\fpsi({\bm r}') \big]  = \big[ \fpsi^{\dagger}({\bm r}), \,
\fpsi^{\dagger}({\bm r}') \big]  =0\;,
\end{equation}
and the coupling constant $g$, for a dilute gas in which the
inter-particle spacing is much larger than the scattering length, is
related to the $s$-wave scattering length $a$ by
\begin{equation}
\label{gcoupling}
g = \frac{4 \pi \hbar^2 a}{m}\;.
\end{equation}

In a zero-temperature BEC, to a very good approximation, all the
atoms are in the ground state $\psi_N({\bm r})$, which is the
$N$-dependent ground-state solution (normalized to unity) of the
time-independent Gross-Pitaevskii (GP) equation for a trapping
potential $V({\bm r})$ and a scattering term with coefficient $g$:
\begin{equation}
\label{eq:timeindepGP}
\left(-\frac{\hbar^2}{2 m } \nabla^2
+V({\bm r})
+g(N-1)|\psi_N|^2\right)\psi_N=\mu_N\psi_N\;.
\end{equation}
Here $\mu_N$ is the chemical potential.  At this level of
approximation, the expansion of the field operator in terms of modal
annihilation operators can be truncated to just one term,
\begin{equation}
\label{creationfield}
\fpsi({\bm r}) = \psi_{N}({\bm r})\hat{a}\;,
\end{equation}
where $\hat a$ annihilates a particle with wave function $\psi_N({\bm
r})$.  The number operator $\hat a^\dagger \hat a$ for this single
mode can be treated as the c-number $N$ because the number of atoms
is a constant.  The Hamiltonian then reduces to the c-number mean-field
energy for this single mode, $H = E_0 N+ \frac{1}{2}g\eta_N N(N-1)$,
where
\begin{equation}
\label{eq:E0}
E_0=\int d{\bm r}\,
\left(\frac{\hbar^2}{2 m }|\nabla \psi_{N}|^2 + V({\bm r}) |\psi_{N}|^2\right)
\end{equation}
is the single-particle kinetic plus trapping energy, and the quantity
\begin{equation}
    \label{eq:eta}
    \eta_N = \int d\bm{r}\,|\psi_{N}(\bm{r})|^4
\end{equation}
is a measure of the inverse volume occupied by the ground-state wave
function.  The product $g\eta_N$, which has units of energy, is a
scattering strength normalized by this effective volume.  The average
number density in the atomic cloud is $N\eta_N$.

So far we have assumed that all the atoms in the BEC are in a single
atomic state, but as mentioned earlier, we want these atoms to be
two-level systems, or qubits, in order for them to serve as the probe
units in the quantum-metrology protocols we are interested in. We
therefore consider two-mode BECs in which the atoms can occupy one of
two internal states, labeled $|1 \rangle$ and $| 2 \rangle$. These
two states are typically hyperfine levels of the atoms.  In practice,
the atoms are cooled to form the BEC while they are all in the same
internal state, and then an external field is used to drive
transitions between the two levels to achieve the desired coherent
superposition of atomic population between the two levels.  The
effect we are looking for is the difference between the integrated
nonlinear phase shifts experienced by the two levels, the difference
being due to the different scattering interactions experienced by the
two levels.  This differential integrated phase shift is detected by
driving a second transition between the levels, which transfers the
phase information into the populations of the two levels.

For an initial analysis of this scenario in this section, we make
three simplifying assumptions:
\begin{enumerate}
\item The external field that drives the transitions between the
two states $|1 \rangle$ and $| 2 \rangle$ acts only for a short
time compared to the phase-shift dynamics that leads to the
estimate of the parameter we are interested in.  We therefore
treat these transitions as effectively instantaneous and do not
include the driving field in the Hamiltonian.

\item The collisions between the atoms are elastic.
Thus the only allowed scattering processes are $|1\rangle | 1
\rangle \rightarrow |1\rangle | 1 \rangle$, $|2\rangle | 2
\rangle \rightarrow |2\rangle | 2 \rangle$, and $|1\rangle | 2
\rangle \rightarrow |1\rangle | 2 \rangle$, with scattering
coefficients $g_{11}$, $g_{22}$, and $g_{12}$, where
$g_{\alpha\beta} = 4 \pi \hbar^2
a_{\alpha\beta}/m=g_{\beta\alpha}$, with Greek letters used to
label the internal states.

\end{enumerate}
These first two assumptions imply that the many-body Hamiltonian
takes the form
\begin{equation}
\label{BEChamil2}
\hat H =
\sum_\alpha\int d{\bm r}\,
\Biggl( \frac{\hbar^2}{2 m } \nabla \fpsi_\alpha^{\dagger}
\cdot \nabla \fpsi_\alpha^{\vphantom{\dagger}}+
V({\bm r})\fpsi_\alpha^{\dagger} \fpsi_\alpha^{\vphantom{\dagger}} \Biggr) +
\frac{1}{2}\sum_{\alpha,\beta}g_{\alpha\beta}
\int d{\bm r}\,\fpsi_\beta^{\dagger} \fpsi_\alpha^{\dagger}
\fpsi_\alpha^{\vphantom{\dagger}} \fpsi_\beta^{\vphantom{\dagger}} \;,
\end{equation}
where $\fpsi_\alpha({\bm r})$ is the field annihilation operator for
internal state~$\alpha$.  In writing this Hamiltonian, we assume that
any energy splitting between the two internal states has been removed
by going to an interation picture.  Our third assumption is by far
the most problematical of the three.
\begin{enumerate}

\item[3.]The two modes retain the same spatial wave function
$\psi_N({\bm r})$ as they evolve.  Since the atoms that form the
initial BEC are all in the state $|1\rangle$, in the mean-field
approximation they all share the spatial wave function
$\psi_{N}({\bm r})$, which is the $N$-dependent ground-state
solution of the time-independent GP
equation~(\ref{eq:timeindepGP}) with scattering coefficient
$g_{11}$. Immediately after the nearly instantaneous action of
the external field, the wave function for both internal states is
$\psi_N({\bm r})$. We further assume that the second internal
state is chosen so that it sees the same trapping potential
$V({\bm r})$.  Even though the two internal states have identical
initial wave functions and experience identical trapping
potentials, their wave functions will gradually become different,
because of the difference in their scattering lengths. What we
are assuming now is that the integrated nonlinear phase shifts
that we are interested in accumulate on a time scale that is
shorter than the time scale for the two wave functions to
differentiate spatially. Thus, for the present, we take the two
wave functions to be identical.  We return to the question of the
time scale for differentiation of the two wave functions at the
end of this section, in Sec.~\ref{subsec:differentiation}.

\end{enumerate}

Using the third assumption, we can write the field annihilation
operators as
\begin{equation}
\fpsi_\alpha({\bm r})=\psi_N({\bm r}) \hat{a}_\alpha\;.
\end{equation}
Since the total number of atoms is fixed, we can treat the total
number operator,
\begin{equation}
\hat{N} \equiv
\hat{a}_1^{\dagger}\hat{a}_1^{\vphantom{\dagger}}+
\hat{a}_2^{\dagger}\hat{a}_2^{\vphantom{\dagger}}\;,
\end{equation}
as a c-number $N$.  We can then put the two-mode Hamiltonian in the
form
\begin{equation}
\label{BEChamil3}
\hat H = E_0 N +
\frac{1}{2}\eta_N\sum_{\alpha,\beta}
g_{\alpha\beta}
\hat{a}_\beta^{\dagger}\hat{a}_\alpha^{\dagger}
\hat{a}_\alpha^{\vphantom{\dagger}}\hat{a}_\beta^{\vphantom{\dagger}}
=H_0 +
\gamma_1\eta_N(N-1)\hat{J}_z+
\gamma_2\eta_N\hat{J}_z^{\,2}\;,
\end{equation}
where $E_0$ and $\eta_N$ are as in Eqs.~(\ref{eq:E0}) and
(\ref{eq:eta}).  The operator $\hat{J}_z$ is defined by
\begin{equation}
\label{eqnj}
\hat{J}_z \equiv \frac{1}{2} \big(
\hat{a}_1^{\dagger}\hat{a}_1^{\vphantom{\dagger}}-
\hat{a}_2^{\dagger}\hat{a}_2^{\vphantom{\dagger}} \big)\;,
\end{equation}
and we have also introduced a c-number energy,
\begin{equation}
H_0 = E_0 N +
\frac{1}{4}\left(\frac{1}{2}(g_{11}+g_{22})+g_{12}\right)\eta_N N^2
-\frac{1}{4}(g_{11}+g_{22})\eta_N N\;,
\end{equation}
which includes the common-mode part of the mean-field scattering
energy. Finally, we define two coupling constants that characterize
the interaction of the two modes,
\begin{equation}
\gamma_1\equiv\frac{1}{2}(g_{11}-g_{22})
\qquad {\mbox{and}} \qquad
\gamma_2\equiv\frac{1}{2}(g_{11}+g_{22})-g_{12}\;.
\end{equation}
The Hamiltonian~(\ref{BEChamil3}) is often called the Josephson
approximation.

The common-mode energy $H_0$ in Eq.~(\ref{BEChamil3}) can be ignored
because its only effect is to introduce an overall phase in the
evolved state of the probe. In the other two terms, we have
$(N-1)\hat{J}_{z}$ and $\hat{J}_{z}^{\,2}$ couplings, suggesting that
we might be able to measure the coupling constants $\gamma_{1}$ and
$\gamma_{2}$ with an accuracy that scales as $1/N^{3/2}$ with the
number of atoms in the BEC.

To see how this works out, suppose the first optical pulse puts each
atom in a superposition $c_1|1\rangle+c_2|2\rangle$, where $c_1$ and
$c_2$ can be assumed to be real (i.e., the first optical pulse
performs a rotation about the $y$ axis of the Bloch sphere). For
short times, we can make a linear approximation to $\hat{J}_z^{\,2}$
in the Josephson Hamiltonian; i.e., we can set
$\hat{J}_z^{\,2}=(\langle \hat{J}_z\rangle+\Delta \hat{J}_z)^2 \simeq
\langle \hat{J}_z\rangle^2+2\langle \hat{J}_z\rangle\Delta
\hat{J}_z$, with $\langle \hat{J}_z\rangle=N(c_1^2-c_2^2)/2$.  The
linear approximation amounts to neglecting the phase dispersion and
corresponding entanglement produced by the $\hat J_z^{\,2}$ term.  We
need not make any such short-time approximation for the $(N-1)\hat
J_z$ term.  Up to irrelevant phases, the resulting evolution is a
rotation of each atom's state about the $z$ axis of the Bloch sphere
with angular velocity
\begin{equation}
\label{eq:OmegaN}
\frac{\eta_N}{\hbar}[(N-1)\gamma_1+N(c_1^2-c_2^2)\gamma_2]
\simeq\frac{(N-1)\eta_N}{\hbar}[\gamma_1+(c_1^2-c_2^2)\gamma_2]\equiv
\Omega_N\;,
\end{equation}
where in the second form, we approximate $N$ as $N-1$.  Under these
circumstances, the BEC acts like a linear Ramsey interferometer whose
rotation rate is enhanced by a factor of $(N-1)\eta_N$, leading to a
sensitivity that scales as $1/\sqrt N (N-1)\eta_N \simeq
1/N^{3/2}\eta_N$.  If $\gamma_2=0$, the optimal initial state has
$c_1=c_2=1/\sqrt2$, but if $\gamma_1=0$, the optimal choice is
$c_1=\cos(\pi/8)$ and $c_2=\sin(\pi/8)$~\cite{boixo08a}.

Achieving a $1/N^{3/2}$ scaling requires that $\eta_N$ have no
dependence on $N$.  As noted above, however, $\eta_N^{-1}$ is a
measure of the volume occupied by the ground-state wave function
$\psi_N$.  As atoms are added to a BEC, the wave function spreads
because of the repulsive scattering of the atoms, thereby reducing
$\eta_N$ as $N$ increases. To pin down how the measurement accuracy
scales with $N$, we need to determine how $\eta_N$ behaves as a
function of $N$.

\subsection{Two critical atom numbers \label{subsec:critical}}

Since we first create a BEC of $N$ atoms all in hyperfine state
$|1\rangle$, before putting them in a superposition of states
$|1\rangle$ and $|2 \rangle$, we can focus on the $N$-dependence of
$\eta_N$ for a single-mode BEC of atoms in state $|1\rangle$. Thus,
in this subsection and the next two, we deal with the single-mode GP
equation~(\ref{eq:timeindepGP}) with $g=g_{11}$ and $a=a_{11}$.

An obvious strategy to suppress the $N$-dependence of $\eta_N$ is to
constrain the BEC within a hard-walled trap so that it cannot expand
as more atoms are added. BECs effectively confined to two or one
dimensions and held in power-law trapping potentials along these
dimensions are the sort found in real experiments. Thus we look at
the dependence of $\eta_N$ on $N$ for a BEC that is {\em loosely\/}
trapped in $d$ dimensions, referred to as {\em longitudinal\/} ($L$)
dimensions, and {\em tightly\/} trapped in $D=3-d$ dimensions,
referred to as {\em transverse\/} (T) dimensions. We assume that in
the longitudinal dimensions, the atoms are trapped in a power-law
potential of the form
\begin{equation}
    \label{eq:traps}
    V_{L}({\bm r}) = \frac{1}{2}kr^{q}\;, \quad q=1,2\ldots,
\end{equation}
and that in the transverse dimensions, the trapping potential is
harmonic,
\begin{equation}
    \label{eq:transtrap}
    V_{T}({\bm \rho}) = \frac{1}{2}m\omega_{T}^{2}\rho^{2}\;.
\end{equation}
The parameter $q$ characterizes the hardness of the longitudinal
trapping potential.  We deal with a 3D trap by setting $D=0$, meaning
there are no transverse dimensions.

When $N$ is small, the mean-field scattering energy is negligible
compared to the atomic kinetic energy of the atoms and the trapping
potential energy.  In this situation, the scattering term in the GP
equation can be neglected, and the ground-state wave function is the
solution of the Schr\"odinger equation for the trapping potential
$V_{L}({\bm r}) + V_{T}({\bm \rho})$.  As more atoms are added to the
BEC, the repulsive scattering term in Eq.~(\ref{eq:timeindepGP})
comes into play and causes the wave function to spread.  We define
two critical atom numbers, $N_{L}$ and $N_{T}$, which characterize
the onset of spreading in the longitudinal and transverse dimensions.
The lower critical atom number, $N_L$, is defined as the atom number
at which the scattering term in the GP equation is as large as the
longitudinal kinetic-energy term and thus characterizes when the wave
function begins to spread in the longitudinal dimensions.  The upper
critical atom number, $N_T$, is defined as the atom number at which
the scattering term is as large as the transverse kinetic energy and
thus characterizes when the wave function begins also to spread in
the transverse dimensions.  The notion of an upper critical atom
number only makes sense for 1D and 2D traps and not for $d=3$.

For small atom number, i.e., $N\ll N_L$, as just noted, the
scattering term in the GP equation can be neglected, and the
ground-state solution of the GP equation is the $N$-independent,
product ground state of the Schr\"odinger equation:
\begin{equation}
    \label{eq:sepwf}
    \psi_0({\bm \rho}, {\bm r}) = \chi_0({\bm \rho})\phi_0({\bm r})\;.
\end{equation}
Here $\chi_0({\bm\rho})$ is the Gaussian ground state for the
transverse dimensions,
\begin{equation}
    \label{eq:transgauss}
    \chi_0({\bm \rho}) =
    \frac{1}{(2 \pi \rho_{0}^{2})^{D/4}}
    \exp \bigg( - \frac{\rho^{2}}{4 \rho_{0}^{2} } \bigg)\;,
\end{equation}
whose corresponding probability density has half-width
\begin{equation}
    \label{eq:rho0}
    \rho_{0} \equiv \bigg( \frac{\hbar}{2 m \omega_{T}}\bigg)^{1/2}\;,
\end{equation}
and $\phi_0({\bm r})$ is the bare ground state for the loosely
confined longitudinal dimensions.   We can estimate the half-width of
$\phi_0$ by equating the trapping potential energy and the kinetic
energy (KE) per dimension, i.e., $kr_0^q/2=\hbar^2/2mr_0^2$, which
gives
\begin{equation}
    \label{eq:r0}
    r_{0} \equiv
    \bigg(\frac{\hbar^{2}}{mk}\bigg)^{1/(q+2)}\;.
\end{equation}
In accordance with our assumptions, we assume that $r_0$ is much
larger than $\rho_0$.  A hard-walled trap in the longitudinal
dimensions corresponds to the limit $q\rightarrow\infty$ with $r_0$
held constant.

The trapped BECs we consider are thus characterized by three length
scales: (i)~the scattering length~$a$; (ii)~the bare transverse trap
half-width $\rho_0$; and (iii)~the bare longitudinal trap half-width
$r_0$.  Typical values, which we use for estimates in the following,
are $a=10\,\mbox{nm}$, $\rho_0=1\,\mu\mbox{m}$, and
$r_0=100\,\mu\mbox{m}$.  For \rbd\ atoms (which have
$a=a_{11}=5.3\,$nm), the corresponding transverse trap frequency is
$\nu_T=58\,$Hz; we can also identify an approximate longitudinal trap
frequency,
\begin{equation}
\label{eq:nuL}
\nu_L=\frac{\omega_L}{2\pi}\equiv\frac{1}{2\pi}\frac{\hbar}{mr_0^2}\simeq10^{-2}\,\mbox{Hz}\;,
\end{equation}
associated with the bare longitudinal ground state.

Whenever the wave function is a product of transverse and
longitudinal wave functions, $\eta_N$ is also a product,
$\eta_N=\eta_T\eta_L$.  When $N\ll N_L$, $\eta_N\equiv\eta_0$
is independent of $N$ since
\begin{eqnarray}
\label{eq:etaT}
\eta_T&=&
\int d^D\!\rho\,|\chi_0({\bm\rho})|^4=\frac{1}{(4\pi)^{D/2}\rho_0^D}\;,\\
\eta_L&=& \int d^d\!r\,|\phi_0({\bm r})|^4 \simeq\frac{1}{V_d r_0^d}\;,
\end{eqnarray}
where $V_d$ is the volume of a unit sphere in $d$ dimensions
($V_1=2$, $V_2=\pi$, and $V_3=4\pi/3$),  The lower critical atom number,
$N_L$, is defined by setting
\begin{equation}
\frac{ \hbar^{2} }{2 m r_{0}^{2}}
\simeq
(\mbox{longitudinal KE})
\simeq
(\mbox{scattering term})
\simeq
(N_L-1)g\eta_0
\simeq
\frac{\hbar^2}{2m}(N_L-1)\frac{1}{\beta_d}\frac{a}{\rho_0^D r_0^d}\;,
\end{equation}
where
\begin{equation}
\beta_d\equiv\frac{V_d}{2(4\pi)^{(d-1)/2}}
\end{equation}
is a geometric factor ($\beta_1=1$, $\beta_2=\sqrt\pi/4$,
$\beta_3=1/6$).  Thus we define
\begin{equation}
    \label{eq:lowern}
    N_L-1\equiv\beta_d
    \frac{r_{0}}{a} \bigg(\frac{\rho_{0}}{r_{0}} \bigg)^{D}\;.
\end{equation}
For the typical length scales mentioned above, the lower critical
atom number is about $1\,700$ for a 3D trap, $45$ for a 2D trap, and
2 for a 1D trap.  The small value of $N_L$ for a 1D trap is the
reason we retain the $-1$ wherever it appears in our discussion of
atom numbers, even though it could be dropped in most situations.

For $N_L\alt N\ll N_T$, the tight confinement in the transverse
dimensions means that the wave function continues to be a product,
\begin{equation}
    \label{eq:sepwfint}
    \psi_{N}({\bm\rho}, {\bm r}) = \chi_0({\bm \rho})\phi_N({\bm r})\;,
\end{equation}
but with the longitudinal wave function satisfying a GP equation,
\begin{equation}
    \label{eq:reducedGP}
    \left(-\frac{\hbar^{2}}{2 m}\nabla_{\!L}^{2}+
    V_{L}({\bm r})+
    g(N-1)\eta_{T}|\phi_N|^{2}\right)\phi_N
    = \mu_{L} \phi_N\;,
\end{equation}
where $\mu_L=\mu_N-D\hbar\omega_T/2$ is the longitudinal part of the
chemical potential.  As atoms are added to the trap in this
intermediate regime, the wave function spreads in the longitudinal
dimensions.  We can estimate the longitudinal half-width $r_N$ by
noticing that $\eta_N=\eta_T\eta_L$, where $\eta_T$ is given by
Eq.~(\ref{eq:etaT}) and
\begin{equation}
\eta_L=\int d{\bm r}\,|\phi_N({\bm r})|^4 \simeq\frac{1}{V_d r_N^d}\;,
\end{equation}
and then equating the attractive longitudinal trapping potential
energy (PE) to the repulsive scattering term:
\begin{eqnarray}
{1\over2}kr_N^q
&\simeq&
(\mbox{longitudinal PE})\nonumber\\
&\simeq&
(\mbox{scattering term})
\simeq
(N-1)g\eta_N
\simeq
\frac{\hbar^2}{2m}(N-1)\frac{1}{\beta_d}\frac{a}{\rho_0^D r_N^d}
=\frac{\hbar^2}{2m}\frac{N-1}{N_L-1}\frac{r_0^{d-2}}{r_N^d}\;,
\end{eqnarray}
where we have used Eq.~(\ref{eq:lowern}) in the last step.  This
leads us to define
\begin{equation} \label{eq:rN}
\frac{r_N}{r_0}\equiv\left(\frac{N-1}{N_L-1}\right)^{1/(d+q)}\;.
\end{equation}
We now define the upper critical atom number by setting
\begin{equation}
\label{eq:upcrit1}
\frac{ \hbar^{2} }{2 m \rho_{0}^{2}}
\simeq
(\mbox{transverse KE})
\simeq
(\mbox{scattering term})
\simeq
(N_T-1)g\eta_{N_T}
=\frac{\hbar^2}{2m}\frac{N_T-1}{N_L-1}\frac{r_0^{d-2}}{r_T^d}\;,
\end{equation}
where $r_T$ is the longitudinal half-width at the upper critical atom
number,
\begin{equation}
\label{eq:upcrit2}
\frac{r_T}{r_0}\equiv\left(\frac{N_T-1}{N_L-1}\right)^{1/(d+q)}\;.
\end{equation}
Using Eq.~(\ref{eq:upcrit1}) and the definitions in
Eqs.~(\ref{eq:upcrit2}) and~(\ref{eq:lowern}) we end up with the
definition
\begin{equation}
\label{eq:upcrit3}
N_T-1\equiv(N_L-1)\left(\frac{r_0}{\rho_0}\right)^{2(d+q)/q}
=\beta_d\frac{\rho_0}{a}
    \bigg(\frac{r_0}{\rho_0}\bigg)^{d(q+2)/q}\;.
\end{equation}
We stress that the notion of an upper critical atom number only makes
sense for 1D and 2D traps and not for $d=3$.  Using the typical
values mentioned above, we have that the upper critical atom number
for a harmonic longitudinal trap ($q=2$) is about $4\times10^9$ for a
2D trap and about $10^6$ for a 1D trap; for a hard longitudinal trap
($q\rightarrow\infty$), $N_T$ is about $4\times10^5$ for a 2D trap
and about $10^4$ for a 1D trap.  Using Eq.~(\ref{eq:upcrit3})
we can rewrite the longitudinal radius in Eq.~(\ref{eq:rN}) as
\begin{equation} \label{eq:rN2}
\frac{r_N}{r_0}=
\left(\frac{r_0}{\rho_0}\right)^{2/q}
\left(\frac{N-1}{N_T-1}\right)^{1/(d+q)}\;.
\end{equation}

It should be noted that
\begin{equation}
\label{eq:rTNT}
\frac{r_T}{\rho_0}=\left(\frac{a}{\rho_0}\frac{N_T-1}{\beta_d}\right)^{1/d}\;.
\end{equation}
For a 1D trap, this gives $r_T=a(N_T-1)$, making the relation between
$r_T$ and $N_T$ independent of the parameters of the trap.  Another
way of thinking about Eq.~(\ref{eq:rTNT}) is that the number density
at the upper critical atom number,
\begin{equation}
\frac{N_T}{\beta_d\rho_0^D r_T^d}\simeq\frac{1}{a\rho_0^2}\;,
\end{equation}
is independent of the properties of the longitudinal trap, with
typical value $10^{14}\,\mbox{cm}^{-3}$.

As the atom number increases from $N_T$, the transverse kinetic
energy becomes unimportant compared to the transverse trapping energy
and the scattering term.  The wave function continues to spread in
the longitudinal dimensions and also spreads in the transverse
dimensions, with the longitudinal and transverse radii, $r_N$ and
$\rho_N$, given by
\begin{eqnarray}
{1\over2}kr_N^q\simeq{1\over2}m\omega_T^2 \rho_N^2
& \simeq &
 (\mbox{scattering term}) \nonumber \\
& \simeq &
(N-1)g\eta_N
\simeq
(N-1)g\frac{1}{V_D\rho_N^D V_dr_N^d}\;,
\end{eqnarray}
which leads us to define in the regime $N \gg N_{T}$,
\begin{eqnarray}
\frac{r_N}{r_0}&\equiv&
\left(\frac{r_0}{2\rho_0}\frac{\rho_N}{\rho_0}\right)^{2/q}\;,\label{eq:rNupper}\\
\left(\frac{\rho_N}{\rho_0}\right)^{5-d+2d/q}
&\equiv&
\frac{4(4\pi)^{D/2}2^{2d/q}}{V_D}\frac{N-1}{N_T-1}\label{eq:rhoN}\;.
\end{eqnarray}

\subsection{Renormalization of the nonlinear interaction terms
and the sensitivity scaling \label{subsec:renormalization}}

The estimates in the previous subsection tell us how $\eta_N$ scales
with atom number.  For atom numbers smaller than the lower critical
atom number, $\eta_N$ has the constant value $\eta_0$, a consequence
of the fact that the repulsive scattering has negligible effect on
the atomic wave function.  In the intermediate regime of atom
numbers, i.e., for atom numbers between $N_L$ and $N_T$, the wave
function expands in the longitudinal dimensions, making $\eta_N$
scale as
\begin{equation}
\label{eq:etaNint}
\eta_N\sim \frac{1}{r_N^d}\sim
\left(\frac{N_L-1}{N-1}\right)^{d/(d+q)}\;.
\end{equation}
For atom numbers above the upper critical atom number, as the wave
function spreads in all dimensions, $\eta_N$ scales as
\begin{equation}
\label{eq:etaNfull}
\eta_N\sim \frac{1}{\rho_N^Dr_N^d}\sim
\frac{1}{\rho_N^{3-d+2d/q}}\sim
\left(\frac{N_T-1}{N-1}\right)^{(3-d+2d/q)/(5-d+2d/q)}\;.
\end{equation}

\begin{figure}[!ht]
\resizebox{7.5 cm}{5 cm}{\includegraphics{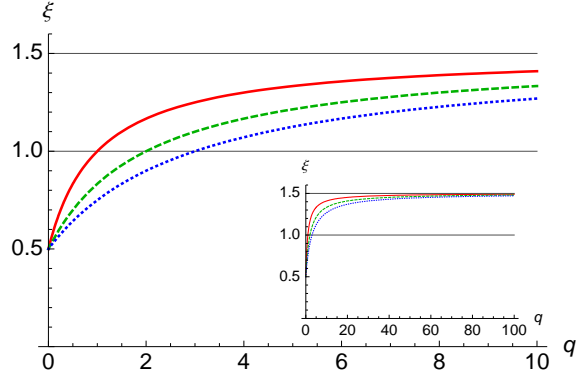}}
\caption{(Color online) Sensitivity scaling exponent
 $\xi=(d+3q)/2(d+q)$ in the intermediate regime of atom numbers,
 $N_L \ll N \ll N_T$, plotted as a function of hardness parameter $q$
 of the longitudinal trapping potential for 1D (red, solid), 2D (green,
 dashed), and 3D (blue, dotted) traps in the intermediate regime of
 atom numbers (for 3D traps, there is no upper critical atom number
 $N_T$).  To achieve super-$1/N$ scaling ($\xi>1$) requires $q>d$.
 A harmonic 1D trap has super-$1/N$ scaling $\xi=7/6$, a harmonic
 2D trap has Heisenberg scaling $\xi=1$, and a 3D harmonic trap has
 sub-$1/N$ scaling $\xi=9/10$.  This sub-$1/N$ scaling for
 3D harmonic traps is still markedly better than the QNL scaling of
 $\xi=1/2$.  A hard-walled trap ($q\rightarrow\infty$) in any
 dimension has $\xi=3/2$.}\label{fig1}
\end{figure}

In the measurement schemes we contemplate, the uncertainties
in determining $\gamma_1$ and $\gamma_2$ scale as
\begin{equation}
\delta\gamma_{1,2}\sim
\frac{1}{\sqrt{N}(N-1)\eta_N}\sim
\frac{1}{N^\xi}\;,
\end{equation}
where in the final form we neglect 1 compared to $N$.  For atom
numbers below $N_L$, the scaling exponent $\xi$ is $3/2$; for
$N_L \ll N \ll N_T$, it takes on the value
\begin{equation} \label{eq:xiintermediate}
\xi=\frac{3}{2}-\frac{d}{d+q}=\frac{d+3q}{2(d+q)}\;;
\end{equation}
and for $N \gg N_T$, $\xi$ is given by
\begin{equation} \label{eq:xifull}
\xi=\frac{3}{2}-\frac{3-d+2d/q}{5-d+2d/q}\;.
\end{equation}
For atom numbers above $N_T$, harmonic 1D and 2D traps have
$\xi=9/10$, a hard-walled 1D trap has $\xi=1$, and a hard-walled 2D
trap has $\xi=7/6$.  Our main interest is the intermediate regime of
Eq.~(\ref{eq:xiintermediate}). The scaling exponent in this regime is
plotted in Fig.~\ref{fig1} as a function of $q$ for 1D, 2D, and 3D
traps.

\subsection{Thomas-Fermi approximations\label{subsec:TF}}

Although we have determined how the scaling exponent behaves with $d$
and $q$, we can do a better job of evaluating $\eta_N$, determining
more precisely the constants in front of the scaling, by using the
Thomas-Fermi (TF) approximation.  In the intermediate regime of atom
numbers, the wave function is the product~(\ref{eq:sepwfint}), with
the longitudinal wave function $\phi_N({\bm r})$ satisfying the GP
equation~(\ref{eq:reducedGP}) in $d$ dimensions.  When $N$ is much
larger than $N_L$, we can ignore the kinetic-energy term in the
reduced GP equation, which gives the TF probability distribution,
\begin{equation}
 \label{eq:intthomasfermi}
 |\phi_N(\bm{r})|^2 = \frac{\mu_{L}-kr^{q}/2}{(N-1)g\eta_{T}} \;.
\end{equation}
Since $ |\phi_N(\bm{r})|^2$ must be positive, the radial extent of
the BEC in the longitudinal dimensions is bounded by $\tilde r_N$
such that
\begin{equation}
\label{Rvalue}
\mu_{L} = \frac{1}{2}k\tilde r_N^q\;.
\end{equation}
Normalization yields
\begin{equation}
\label{norm}
1 = \int d^d\!r\,|\phi_N(\bm{r})|^2\equiv I_1(N,d,q)\;.
\end{equation}
Integrals over TF probability distributions, such as
$I_1$, are defined and evaluated in an Appendix.  Using
Eq.~(\ref{eq:I1}), we find
\begin{equation}
\frac{\tilde r_N}{r_0}=\left(\frac{d+q}{q}\frac{N-1}{N_L-1}\right)^{1/(d+q)}
\end{equation}
[cf.~Eq.~(\ref{eq:rN})] or, equivalently, using Eqs.~(\ref{eq:A1}) and
(\ref{eq:A4}),
\begin{equation}
\frac{\mu_L}{(N-1)g\eta_T}=
\frac{d+q}{q}\frac{1}{V_d\tilde r_N^d}=
\frac{1}{V_d r_0^d}
\left(\frac{d+q}{q}\right)^{q/(d+q)}
\left(\frac{N_L-1}{N-1}\right)^{d/(d+q)}\;.
\end{equation}

Now, from Eq.~(\ref{eq:Il}), we can find
\begin{equation}
\label{etaL}
\eta_{L} = I_2(N,d,q)=
\frac{2q}{d+2q}
\frac{\mu_L}{(N-1)g\eta_{T}}=
\frac{1}{V_dr_0^d}\frac{2q}{d+2q}
\left(\frac{d+q}{q}\right)^{q/(d+q)}
\left(\frac{N_L-1}{N-1}\right)^{d/(d+q)}
\end{equation}
and thus determine $\eta_N=\eta_T\eta_L$
[cf.~Eq.~(\ref{eq:etaNint})].  Numerical computation of $\eta_N$ in
the intermediate regime indicates that this expression is quite
accurate in spite of the approximations that went into obtaining it.

When $N$ is much larger than $N_T$, we can again use a TF
approximation, this time ignoring the kinetic-energy term in the 3D
GP equation~(\ref{eq:timeindepGP}), which gives the probability
distribution
\begin{equation}
 \label{eq:upperthomasfermi}
 |\psi_N(\bm\rho,\bm r)|^2 =
 \frac{\mu_N-m\omega_T^2\rho^2/2-kr^{q}/2}{(N-1)g} \;.
\end{equation}
Positivity of this distribution requires that $\rho\le\tilde\rho_N$
and $r\le\tilde r_N(\rho)$, where
\begin{eqnarray}
\frac{1}{2}m\omega_T^2\tilde\rho_N^2 &=& \mu_N\;,\\
\frac{1}{2}k\tilde r_N^q(\rho)&=&\frac{1}{2}m\omega_T^2(\tilde\rho_N^2-\rho^2)\;.
\end{eqnarray}
The extent of the atomic cloud in the longitudinal direction is
characterized by $\tilde r_N\equiv\tilde r_N(0)$, i.e., $\tilde
r_N/r_0=[(r_0/2\rho_0)(\tilde\rho_N/\rho_0)]^{2/q}$. Normalization
yields
\begin{equation}
1=\int d^D\!\rho\,d^d\!r\,|\psi_N(\bm\rho,\bm r)|^{2}\equiv
K_1(N,d,q)\;.
\end{equation}
Using Eq.~(\ref{eq:K1}), we find
\begin{equation}
\left(\frac{\tilde\rho_N}{\rho_0}\right)^{5-d+2d/q}
=\frac{1}{dJ_{1+d/q}(D,2)J_1(d,q)}
\frac{4(4\pi)^{D/2}2^{2d/q}}{S_{\!D-1}}
\frac{N-1}{N_T-1}
\end{equation}
[cf.~Eq.~(\ref{eq:rhoN})].  We can also write
\begin{equation}
\frac{\mu_N}{(N-1)g}=
\frac{1}{4(4\pi)^{D/2}\rho_0^D V_d r_0^d}
\left(\frac{\rho_0}{r_0}\right)^{2q/d}
\left(\frac{\tilde\rho_N}{\rho_0}\right)^2
\frac{N_T-1}{N-1}
\end{equation}
and
\begin{equation}
\eta_N\equiv K_2(N,d,q)=\frac{2(d+q)}{(D/2+d/q+1)(d+2q)}\frac{\mu_N}{(N-1)g}\;.
\end{equation}
The scaling of $\eta_N$ agrees with that in Eq.~(\ref{eq:etaNfull}).

\subsection{Bose-condensed \rbd~atoms}
\label{subsec:rubidium}

A good candidate for implementing the generalized metrology protocol
is a BEC made of Rubidium (\rbd) atoms~\cite{williams99a,matthews98a,
hall98a,hall98b}.  Initially the atoms in the condensate are in the
ground electronic state.  In most experiments~\cite{matthews98a,
hall98a,hall98b}, the $|F=1;\,M_F=-1\rangle\equiv|1\rangle$ state is
trapped and cooled to the condensation point.  Once the atoms in
$|1\rangle$ have accumulated in the condensate ground state, a
two-photon drive is used to couple the $|1\rangle$ state to the
$|F=2;\,M_F=+1\rangle\equiv|2\rangle$ state. The two-photon drive
involves applying both microwave and radio-frequency electromagnetic
fields to the condensate. Both hyperfine states are not cooled
simultaneously to form a condensate in a superposition of the two
states because the lifetime of atoms in the $|2\rangle$ state in a
trap is much shorter than the lifetime of atoms in the $|1\rangle$
state. As mentioned earlier, we assume that the driving field that
initializes the atoms in a desired superposition of the $|1\rangle$
and $|2\rangle$ states is instantaneous in comparison to the dynamics
that is part of the parameter-estimation process.

The s-wave scattering lengths for the three processes, $|1 \rangle |
1 \rangle \rightarrow |1 \rangle | 1 \rangle$, $|2 \rangle | 2
\rangle \rightarrow |2 \rangle | 2 \rangle$, $|1 \rangle | 2 \rangle
\rightarrow |1 \rangle | 2 \rangle$ are  nearly degenerate for \rbd,
with the ratios $\{ a_{22} : a_{12} : a_{11} \} = \{0.97 : 1 :
1.03\}$ ($a_{11}=5.31\,$nm)~\cite{williams99a}. These values for the
scattering lengths mean that $\gamma_2 = (g_{11}+g_{22})/2-g_{12} =0$
for \rbd.  Therefore, the \rbd~BEC realizes the generalized
quantum-metrology protocol with just the $\gamma_1(N-1)\eta_N
\hat{J}_z$ coupling in the Hamiltonian~(\ref{BEChamil3}).

The optimal initial state for metrology with the
$\gamma_{1}(N-1)\eta_N\hat{J}_{z}$ coupling is the one in which all
atoms are initialized in the equatorial plane of the Bloch sphere,
say, in the $+1$ eigenstate of $\sigma_x$,
$(|1\rangle+|2\rangle)/\sqrt{2}$. The quantity estimated by this
measurement scheme is $\gamma_1=(g_{11}-g_{22})/2$, which is small,
but finite in the case of a \rbd~BEC.  Once the probe is initialized
in the optimal initial state, we let it evolve for a time $t$ under
the influence of the $\gamma_1(N-1)\eta_N\hat{J}_z$ Hamiltonian,
which simply rotates the state of each atom about the $z$ axis of the
Bloch sphere with angular velocity $\gamma_1(N-1)\eta_N/\hbar$.  At
the end of this evolution, we measure an equatorial component of
$\hat{\bm J}$ ($\hat{J}_y$ for short evolution times), which is
achieved by a $\pi/2$ pulse about the desired equatorial axis,
followed by a measurement of $\hat J_z$, i.e., of the difference in
the populations of the two internal states.

Precision experiments with two-component \rbd~BECs in modestly
nonspherical, harmonic traps have been reported in \cite{mertes07a}
and \cite{anderson09}.  These experiments were carried out with atom
numbers in excess of $100\,000$ and thus operated in the full TF
regime well above the upper critical atom number.

\subsection{Differentiation of the spatial wave functions
\label{subsec:differentiation}}

The strongest assumption we made in obtaining the
Hamiltonian~(\ref{BEChamil3}) was that the wave functions for the two
modes remain identical throughout the duration of the proposed
measurement scheme.  Here we examine this assumption more carefully.

We noted earlier that even if the two modes see the same trapping
potential, the difference in their scattering lengths will cause the
two wave functions to evolve differently~\cite{matthews98a}.
The initial effect of the difference in scattering lengths is to
produce a relative phase between $|1\rangle$ and $|2\rangle$.  This
relative phase depends on the local density within the condensate.
The integrated (or average) part of the relative phase provides the
signal for our measurement protocol, whereas the residual
position-dependent part of the relative phase reduces the visibility
of the fringes on which the signal relies.  For our protocol to
succeed, we need the integrated phase to accumulate more rapidly than
the residual position-dependent phase.  Yet a further effect is that
the position-dependent phases drive differences between the atomic
densities associated with the two hyperfine
levels, but as this occurs on a longer time scale than the
accumulation of the position-dependent phase shift, we do not
consider it here.

We can analyze this scenario in the following way.  Initially all
atoms are in the state $\psi_N({\bm\rho},{\bm r})|1\rangle$.  After
the first optical pulse, the state becomes $\psi_N({\bm\rho},{\bm
r})(c_1|1\rangle+c_2|2\rangle)$, where $c_{1}$ and $c_{2}$ are the
amplitudes to be in the hyperfine states.  We can assume that $c_1$
and $c_2$ are real, i.e., that the initial optical pulse produces a
rotation about the $y$ axis of the Bloch sphere.  The different
scattering lengths make the wave functions for the two modes evolve
differently, so that after a time~$t$, the atomic state becomes
$c_1\psi_{N,1}({\bm\rho},{\bm r},t)|1\rangle+
c_2\psi_{N,2}({\bm\rho},{\bm r},t)|2\rangle$,
where the wave functions for the two modes evolve according to
time-dependent, coupled GP equations:
\begin{equation}
\label{eq:coupledGP}
i\hbar \frac{\partial\psi_{N,\alpha}}{\partial t} =
        \left(-\frac{\hbar^2}{2m}\nabla^2+V+
        (N-1)\sum_{\beta}g_{\alpha\beta}c_\beta^2|\psi_{N,\beta}|^2\right)
        \psi_{N,\alpha}\;.
\end{equation}
The second optical pulse is a $\pi/2$ pulse about an equatorial axis
of the Bloch sphere.  For the discussion here, we assume that this
rotation is about the $x$ axis so that subsequent counting of the
populations of the two hyperfine levels is equivalent to measuring
$\hat{J}_y$ before the second optical pulse. The state after the
pulse is
\begin{equation}
    \label{eq:GPfinalstate}
    c_1\psi_{N,1}\frac{1}{\sqrt2}\bigl(|1\rangle-i|2\rangle\bigr)
    +c_2\psi_{N,2}\frac{1}{\sqrt2}\bigl(-i|1\rangle+|2\rangle\bigr)=
    \frac{1}{\sqrt2}\bigl(c_1\psi_{N,1}-ic_2\psi_{N,2}\bigr)|1\rangle-
    \frac{i}{\sqrt2}\bigl(c_1\psi_{N,1}+ic_2\psi_{N,2}\bigr)|2\rangle\;.
\end{equation}
The corresponding probabilities to be in the two states,
\begin{equation}
    \label{eq:probs}
    p_{1,2}=\frac{1}{2}\bigl[1\mp
    2c_1c_2\mbox{Im}(\langle\psi_{N,2}|\psi_{N,1}\rangle)\bigr]\;,
\end{equation}
are determined by the overlap of the two spatial wave functions,
\begin{equation}
\langle\psi_{N,2}|\psi_{N,1}\rangle=
\int d^D\!\rho\,d^d\!r\,\psi_{N,2}^*\psi_{N,1}\;.
\end{equation}

In a ground-breaking set of experiments, Anderson~{\em et~al.}
\cite{anderson09} measured the position-dependent phase shifts in a
two-component \rbd~BEC, trapped in a modestly nonspherical trap, and
saw the associated reduction in fringe visibility. The details of the
experiment were shown to be well accounted for by numerical
integrations of the two-component GP equations~(\ref{eq:coupledGP})
with a loss term included.  The experiment was carried out with atom
number $N\simeq 1.5\times10^5$, well above the upper critical atom
number.

To compare the time scales for the integrated and position-dependent
phase shifts in our protocol, we assume that we are operating in the
intermediate regime of atom numbers, i.e., $N_L\alt N\ll N_T$.  In
this regime, the wave functions for the two modes factor into
transverse and longitudinal wave functions, i.e.,
\begin{equation}
\psi_{N,\alpha}({\bm\rho},{\bm r},t)=
\chi_0({\bm\rho})\phi_{N,\alpha}({\bm r},t), \quad \alpha =1,2\;,
\end{equation}
where $\chi_0$ is the time-independent, Gaussian ground state in the
transverse dimensions and the longitudinal wave functions obey
time-dependent, coupled, longitudinal GP equations,
\begin{equation}
\label{eq:coupledGPlong}
i \hbar \frac{\partial\phi_{N,\alpha}}{\partial t} =
        \left(-\frac{\hbar^2}{2m}\nabla_L^2+V_{L}+
        \eta_T(N-1)\sum_{\beta}g_{\alpha\beta}c_\beta^2|\phi_{N,\beta}|^2\right)
        \phi_{N,\alpha}\;.
\end{equation}

To estimate the time scales, we assume that $N$ is large enough
relative to the lower critical atom number to justify the TF
approximation in the longitudinal dimensions, thus allowing us to
ignore the kinetic-energy terms in the coupled GP equations.  With
these assumptions, the probability densities do not change with time,
i.e.,
\begin{equation}
|\phi_{N,\alpha}({\bm r},t)|^2=|\phi_N({\bm r},0)|^2\equiv q_0({\bm r})\;,
\end{equation}
and the evolution under the coupled GP equations only introduces a phase,
\begin{equation}
    \label{eq:phitrans}
    \phi_{N,\alpha}({\bm r}, t) =
    \sqrt{q_0}\exp\bigg[-\frac{it}{\hbar}\bigg(V_L+
    \eta_T(N-1)q_0
    \sum_{\beta} g_{\alpha \beta}c_{\beta}^{2}\bigg) \bigg]\;.
\end{equation}
This gives an overlap $\langle\psi_{N,2}|\psi_{N,1}\rangle
=\langle\phi_{N,2}|\phi_{N,1}\rangle =\int d^d\!r\,
q_0e^{-i\delta\theta({\bm r})}$, where the relative phase
is given by
\begin{equation}
\label{eq:deltatheta}
\delta\theta({\bm r})=\frac{\eta_T(N-1)q_0({\bm r})\Delta g\,t}{\hbar}=
\Omega_N t\left(1+\frac{q_0({\bm r})-\eta_L}{\eta_L}\right)\;,
\end{equation}
with
\begin{equation}
\Delta g\equiv c_{1}^{2}(g_{11}-g_{12})-c_{2}^{2}(g_{22}-g_{12})
=\gamma_{1} + (c_1^2-c_2^2)\gamma_{2}\;.
\end{equation}
In the second equality of Eq.~(\ref{eq:deltatheta}), we have
separated out the integrated phase shift, which has angular frequency
\begin{equation}
\Omega_N\equiv\frac{(N-1)\eta_N\Delta g}{\hbar}=
\omega_L\frac{\Delta g}{g_{11}}\frac{q}{d+2q}
\left(\frac{q}{d+q}\frac{N-1}{N_L-1}\right)^{q/(d+q)}
\end{equation}
[cf.~Eqs.~(\ref{eq:OmegaN} and~(\ref{eq:nuL})], leaving the residual
position-dependent phase shift as a correction.  The final expression
for $\Omega_N$ uses the TF approximation to evaluate $\eta_N$ in the
intermediate regime.  For the \rbd~protocol outlined in
Sec.~\ref{subsec:rubidium}, in which $\gamma_2$ is essentially zero,
we choose $c_1^2=c_2^2=1/2$ in order to maximize the fringe
visibility in Eq.~(\ref{eq:probs}).

It is worth emphasizing how this approach based on coupled GP
equations differs from use of the Josephson
Hamiltonian~(\ref{BEChamil3}).  Although the GP equations yield a
position-dependent phase, which cannot be obtained from the Josephson
Hamiltonian, this comes at a price: the integrated relative phase in
Eq.~(\ref{eq:deltatheta}) amounts to making the linear approximation
to $\hat{J}_z^{\,2}$ described in the paragraph containing
Eq.~(\ref{eq:OmegaN}).  The linear approximation is essential because
the $\hat{J}_z^{\,2}$ coupling does not preserve product states,
whereas the GP equations assume a product state.  It means that the
GP equations miss the phase dispersion generated by the $\hat
J_z^{\,2}$ coupling and the associated dynamically generated
entanglement.

We can now write the overlap as
\begin{equation}
    \label{eq:theoreticaloverlap}
\langle\psi_{N,2}|\psi_{N,1}\rangle=
e^{-i\Omega_N t}
\int d^d\!r\,q_0
e^{-i\Omega_N t(q_0-\eta_L)/\eta_L}
\simeq
e^{-i\Omega_N t}
\exp\left(-\frac{\Omega_N^2 t^2}{2\eta_L^2}
\int d^d\!r\,q_0(q_0-\eta_L)^2\right)\;,
\end{equation}
where the second expression comes from expanding the exponential
inside the integral to second order and then converting to an
equivalent Gaussian at the same order.  The contribution from the
first-order term vanishes since $\eta_{L}=\int d^d\!r\,q_{0}^{2}$. We
can identify a time scale $\tau_{\rm pd}$ for the position-dependent
phase as the time set by the half-width of the Gaussian, i.e.,
\begin{equation}
\label{eq:Omegatau}
\Omega_N\tau_{\rm pd}\equiv\eta_L\left(\int d^d\!r\,q_0(q_0-\eta_L)^2\right)^{-1/2}
=\sqrt{\frac{2(d+3q)}{d}}\;.
\end{equation}
The final form comes from using the TF
approximation~(\ref{eq:intthomasfermi}) for the density $q_0$ and the
results in the Appendix to evaluate the integral.

What this result means is that to retain good fringe visibility, our
protocol will generally be restricted to operating well within the
first fringe.  One can expect, however, that as the longitudinal trap
becomes more hard-walled, the TF density becomes more and more
flat-topped, eventually approaching a box, with the result that the
residual position-dependent phase shift becomes smaller and smaller.
This expectation is borne out by Eq.~(\ref{eq:Omegatau}), which
reports that $\tau_{\rm pd}$ gets larger as the hardness parameter
$q$ increases; e.g., for a 1D trap with $q=10$, $\Omega_N\tau_{\rm
pd}\simeq 8$.

To investigate further this way of reducing the effect of the
position-dependent phase requires numerical simulations and more
accurate approximation procedures, both of which we have undertaken.
Initial results, to be reported elsewhere, suggest that things turn
out better than is suggested by the crude approximations that go into
Eq.~(\ref{eq:Omegatau}).

\subsection{Other practical considerations \label{subsec:practical}}

\subsubsection{Loss of Atoms \label{subsubsec:loss}}

For a one-component BEC, the total number of atoms for a given trap
is limited by three-body losses.  This process is usually the most
significant loss channel, with all other losses being negligible. For
a two-component BEC, however, things are different because other loss
channels, such as inelastic two-atom (spin-exchange) collisions,
become significant even when the number of atoms in the trap is such
that three-body collisions are unimportant.  Just as in the case of
three-body collisions, the spin-exchange collisions can be considered
as a process that leads to loss of atoms from the trap.

Spin-exchange collisions in a two-component BEC of \rbd~atoms in the
two hyperfine levels we are interested in were considered
in~\cite{mertes07a} and \cite{anderson09}.  The effect of inelastic
spin-exchange interactions was modeled by including non-Hermitian
potentials in the coupled GP equations~(\ref{eq:coupledGP}):
\begin{eqnarray}
&&-\frac{i\hbar}{2}(N-1)\Gamma_{12}c_2^2|\psi_2|^2\quad\mbox{for mode~1,}\\
&&-\frac{i\hbar}{2}(N-1)\Bigl(\Gamma_{12}c_1^2|\psi_1|^2+\Gamma_{22}c_2^2|\psi_2|^2\Bigr)
\quad\mbox{for mode~2.}
\end{eqnarray}
The loss constants in \rbd\ were measured to be
$\Gamma_{12}=0.780(19)\times 10^{-13}\,\mathrm{cm^3/s}$ and
$\Gamma_{22}=1.194(19)\times 10^{-13}\,\mathrm{cm^3/s}$.  If we
assume that the wavefunctions are the same for the two hyperfine
states, as in the short-time analysis of
Sec.~\ref{subsec:differentiation}, the integrated effect of the
spin-exchange losses across the atomic cloud is characterized by a
decay constant
\begin{equation}
\Gamma\equiv\frac{(N-1)\eta_N(\Gamma_{12}+\Gamma_{22}c_2^2)}{2}\;.
\end{equation}
We can get an idea of the importance of spin-exchange losses by
comparing $\Gamma$ to the angular frequency $\Omega_N$ for the
integrated phase shift.  The ratio of interest for comparing coherent
and decoherent processes is thus
\begin{equation}
\label{eq:decratio}
\frac{\Gamma}{\Omega}_N=
\frac{\hbar(\Gamma_{12}+\Gamma_{22}/2)}{2\gamma_1}=
\frac{m}{4\pi\hbar}\frac{\Gamma_{12}+\Gamma_{22}/2}{a_{11}-a_{22}}
\simeq\frac{1}{19}\;,
\end{equation}
where we specialize to the case $c_1^2=c_2^2=1/2$ relevant to the
\rbd~protocol and the final estimate applies to that protocol. This
ratio indicates that the proposed protocol can obtain an estimate of
$\gamma_{1}$ with better than $1/N$ scaling before atom losses
degrade the sensitivity.

It is also to be noted that an advantage of using a measurement
scheme that uses product states is that loss of atoms from the BEC
does not change the sensitivity scaling, since loss of particles from
a product state does not damage any coherence.  There is a decay in
the signal strength given by a factor $e^{-\Gamma t}$, which would
require us to complete the experiment before too many atoms are lost,
but the ratio~(\ref{eq:decratio}) provides a window for doing this.
The discussion in Sec.~\ref{subsec:differentiation} suggests,
however, that differentiation of the spatial wave functions for the
two modes becomes a limiting factor on the duration of the experiment
before loss of atoms becomes an important consideration.

\subsubsection{Number uncertainties \label{subsubsec:number}}

In real experiments the number of atoms in a BEC is not known to
arbitrary precision as we have assumed so far.  Thus we have to
consider what happens when the number of atoms in the BEC is not
fixed from trial to trial.

To analyze this situation, let $p(N_0)$ denote the probability that
the number of atoms participating in our measurement protocol is
$N_0$.  The final step in the protocol is to count the number of
atoms in the two hyperfine levels.  The difference between the two
counts is used to estimate the parameter, here denoted as $\gamma$;
the sum can be used to refine the estimate of the number of atoms
that participated in the protocol.

We let $N'_1$ and $N'_2$ be the number of atoms that would be counted
by an ideal counting procedure.  We generally work in terms of the
total number of atoms, $N_0=N'_1+N'_2$, and the difference,
$m'=(N'_1-N'_2)/2$, normalized by a factor of two to match the
eigenvalues of $\hat J_z$.  Quantum mechanics gives the conditional
probability $q(m'|N_0,\gamma)$ for a measurement of $\hat J_z$.

The counting is not completely precise, so we introduce independent
conditional probabilities, $p(N_1|N'_1)$ and $p(N_2|N'_2)$, for
counting $N_1$ and $N_2$ atoms in the two levels, given the ideal
counts.  We can think of these two probabilities as describing
processes in which condensate atoms are missed or non-condensate
atoms are counted by mistake.  In addition, in a complete analysis of
the protocol, we would need to include the loss of atoms, discussed
in the previous subsection, in this analysis.  As already noted, we
are mainly interested in the total number of atoms counted,
$N=N_1+N_2$, and the normalized difference, $m=(N_1-N_2)/2$.  In the
absence of a better model, we assume, to illustrate the effect of
number uncertainties, that $p(N_1|N'_1)$ and $p(N_2|N'_2)$ are
independent Gaussian random processes, with mean
$\overline{N_j}=N'_j$ and variance $\Delta^2 N_j=\sigma^2$. Under
this assumption, $N$ and $m$ become independent Gaussian random
processes, described by conditional probabilities $p(N|N_0)$ and
$p(m|m')$, which have $\overline{N}=N_0$, $\overline{m}=m'$,
$\Delta^2 N=2\sigma^2$, and $\Delta^2 m=\sigma^2/2$.

The probability this model gives us directly is
\begin{equation}
p(N,m,m',N_0|\gamma)=p(N|N_0)p(m|m')q(m'|N_0,\gamma)p(N_0)\;,
\end{equation}
The probability we need in order to evaluate the sensitivity of our
protocol is the conditional probability for $m$, given the parameter
$\gamma$ and the measured total number of atoms, $N$:
\begin{eqnarray}
\label{eq:probmprime}
p(m|N,\gamma)&=&\frac{p(N,m|\gamma)}{p(N|\gamma)}\nonumber\\
&=&\frac{\displaystyle{\sum_{m',N_0}p(N,m,m',N_0|\gamma)}}
{\displaystyle{\sum_{m,m',N_0}p(N,m,m',N_0|\gamma)}}\nonumber\\
&=&\sum_{m',N_0}p(m|m')q(m'|N_0,\gamma)p(N_0|N)\;.
\end{eqnarray}
In the final form, $p(N_0|N)=p(N|N_0)p(N_0)/p(N)$ is the
conditional probability for $N_0$ atoms to have participated in the
protocol, given the measured total count $N$.  It quantifies the
refinement in the knowledge of $N_0$ provided by the total count.

The quantities that go into determining the sensitivity are the mean
and second moment of $m$, calculated from the
probability~(\ref{eq:probmprime}),
\begin{eqnarray}
\overline{m}_{N,\gamma}&=&\sum_{N_0}\langle\hat J_z\rangle_{N_0,\gamma}p(N_0|N)\;,\\
(\overline{m^2})_{N,\gamma}
&=&\frac{1}{2}\sigma^2+\sum_{N_0}\Bigl(\langle\hat J_z\rangle^2_{N_0,\gamma}+
(\Delta^2\hat J_z)_{N_0,\gamma}\Bigr)p(N_0|N)\;,
\end{eqnarray}
where $\langle\hat
J_z\rangle_{N_0,\gamma}=(\overline{m'})_{N_0,\gamma}$ and
$(\Delta^2\hat J_z)_{N_0,\gamma}=(\Delta^2 m')_{N_0,\gamma}$ are the
mean and variance of $\hat J_z$ calculated from the
quantum-mechanical probabilities.  If $\sigma$ is much less than the
initial uncertainty in $N_0$, which is itself somewhat less than
$N_0$ (depending on the care taken in loading the trap), then the
measured total count $N$ gives a very good, improved estimate of the
number of atoms that participated in the protocol; under these
circumstances, the probability $p(N_0|N)$ is peaked at the measured
value $N$, with half-width given very nearly by $\sigma$.  The
quantum-mechanical expectation values vary over a range from $-N_0/2$
to $+N_0/2$, so as long as $\sigma\ll N$, we can evaluate the
averages over $p(N_0|N)$ at the mean value $N$ with little error,
thus giving mean $\overline{m}_{N,\gamma}=\langle\hat
J_z\rangle_{N,\gamma}$ and variance $(\Delta^2
m)_{N,\gamma}=\sigma^2/2+(\Delta^2\hat J_z)_{N,\gamma}$.  The
resulting measurement uncertainty in determining $\gamma$,
\begin{equation}
    \label{eq:uncertn}
    \delta \gamma^2 =
    \frac{(\Delta^2 m)_{N,\gamma}}
    {|\partial\overline{m}_{N,\gamma}/\partial\gamma|^{2}}
    =\frac{\sigma^2/2+(\Delta^2\hat J_z)_{N,\gamma}}
    {|\partial\langle\hat J_z\rangle_{N,\gamma}/\partial\gamma|^{2}}\;,
\end{equation}
has the quantum-mechanical scaling and nearly the optimal
sensitivity, provided we can count atoms to better than $\sqrt N$,
i.e., $\sigma\alt\sqrt N$. Ultimately, what this result expresses is
that the variance of the measurement of $\hat J_z$ in our protocol is
of order $\sqrt N$, so we need to know the number of atoms to this
same accuracy.

\section{Conclusion \label{sec:conclusion}}

This paper serves two purposes.  The first is to extend the
discussion of Heisenberg-limited quantum metrology from its
traditional focus on a $1/N$ scaling for measurement uncertainty. Our
discussion centers on the role of the dynamics of an $N$-qubit
quantum probe in determining the quantum \crb\ for single-parameter
estimation.  Looking at quantum metrology using the language of
quantum circuits makes it easy to see that abandoning the usual
independent couplings of the parameter to the qubits in favor of
nonlinear couplings can yield scalings better than $1/N$.  With
$k$-body couplings, it is possible to achieve sensitivities scaling
as $1/N^k$.  Although the $1/N^k$ scaling requires entangled input
states, $1/N^{k-1/2}$ scalings can be obtained with initial product
states.  Thus a sensitivity scaling as $1/N^{3/2}$ can be achieved if
quadratic couplings to the parameter can be engineered; moreover,
particular quadratic couplings yield this sensitivity even though the
state remains unentangled under the dynamics, thus showing that
super-$1/N$ scaling can be achieved without any entanglement.

The second purpose of this paper is to show that a two-component BEC
is a promising candidate system for a proof-of-principle experiment
that demonstrates scaling better than $1/N$.  A simplified analysis
of the system, based on strong assumptions, but followed by a more
detailed analysis of the realm of applicability of those assumptions,
shows that such an experiment might indeed be realizable. This work
motivates further, yet more detailed analyses and numerical
simulations of the experiment.  We have undertaken such further
investigations of the proposed metrology scheme, and this further
work, to be reported elsewhere, supports the conclusions reached in
this paper.  Our numerical studies include computing the ground-state
solution of the time-independent GP equation for different values of
$N$ in order to find the exact dependence of $\eta_{N}$ on $N$.
Numerical integration of the time-depenent, coupled, two-mode GP
equations~(\ref{eq:coupledGP}) is then used to compute the expected
signal~(\ref{eq:probs}) in order to compare it with the theoretical
prediction in Eq.~(\ref{eq:theoreticaloverlap}).

The quantity that is measured in the proposed metrology protocol is
essentially a constant.  Estimating a constant using sophisticated
quantum measurement schemes is interesting only as a proof of
principle, because there is nothing to preclude estimating the same
constant using much simpler, classical measurement techniques. Since
the measured quantity is a constant, we have the time to perform
whatever number of repetitions of the simplest estimation procedure
is required to achieve the desired accuracy. Metrology protocols of
the type described here are relevant in circumstances where there are
constraints on the available time or on the available number of
qubits.  The available time can be constrained, for example, because
the quantity that is being measured is changing, as in the case of
gravitational-wave detection or magnetometry.  There can be further
time constraints placed by decoherence of the probe qubits.  In such
scenarios, picking the optimal metrology scheme with the best
measurement uncertainty, given the constraints, is of primary
importance~\cite{shaji07a}.  For our proposal using a BEC, one
possibility is to work around a broad Feshbach resonance that makes
the scattering lengths sensitive to external magnetic fields. We
might then be able to use our scheme for high-precision magnetometry.

\acknowledgements

This work was supported in part by the US Office of Naval Research
(Grant No.~N00014-07-1-0304) and the Australian Research Council's
Discovery Projects funding scheme (Project Nos.~DP0343094 and
DP0985142). SB was supported by the National Science Foundation under
grant PHY-0803371 through the Institute for Quantum Information at
the California Institute of Technology.  AD was supported in part by
the EPSRC (Grant No. EP/C546237/1), EPSRC QIP-IRC, EU Integrated Project
(QAP) and the EU STREP project HIP.

\appendix*

\section{Integrals over Thomas-Fermi distributions}

In the intermediate TF regime, i.e., $N_L\ll N\ll N_T$, we need to do
integrals over the TF probability density~(\ref{eq:intthomasfermi}),
\begin{eqnarray}
\label{eq:A1}
I_l(N,d,q)&\equiv&
\int d^d\!r\,|\phi_N(\bm{r})|^{2l} \nonumber \\
& = & \left(\frac{k/2}{(N-1)g\eta_{T}}\right)^l
\int d\Omega_{d-1} \int_0^{\tilde r_N} r^{d-1}dr\,(\tilde r_N^q - r^q)^l \nonumber \\
& = & \left(\frac{k/2}{(N-1)g\eta_{T}}\right)^l \tilde r_N^{d+ql} S_{\!d-1}
J_l(d,q) \nonumber \\
& = & \left(\frac{\mu_L}{(N-1)g\eta_{T}}\right)^l
\tilde r_N^d S_{\!d-1} J_l(d,q)\;,
\end{eqnarray}
where $S_{\!d-1}=dV_d$ is the area of a unit sphere in $d-1$
dimensions and
\begin{equation}
\label{eq:A2}
J_l(d,q)\equiv
\int_0^1 du\,u^{d-1}(1-u^q)^l=
-\frac{\pi}{\sin(l\pi)\Gamma(-l)}\frac{\Gamma(d/q)}{q\Gamma(d/q+l+1)}
\end{equation}
for $l>-1$.

It is easy to see that
\begin{equation}
\label{eq:A3}
\frac{J_{x+l}(d,q)}{J_x(d,q)}=\frac{(x+1)\cdots(x+l)}{(d/q+x+1)\cdots(d/q+x+l)}\;.
\end{equation}
Combined with $J_0=1/d$, this gives, when $l$ is a nonnegative
integer,
\begin{equation}
\label{eq:A4}
J_l(d,q)=\frac{l!\,q^l}{d(d+q)(d+2q)\cdots(d+lq)}\;.
\end{equation}
Notice also that
\begin{equation}
\label{eq:A5}
J_l(d,2)=\int_0^{\pi/2}dv\,\sin^{d-1}\!v\cos^{2l+1}\!v=
\frac{\Gamma(d/2)\Gamma(l+1)}{2\Gamma(d/2+l+1)}\;.
\end{equation}

Now we use
\begin{equation}
\label{eq:A6}
\frac{k/2}{(N-1)g\eta_T}S_{\!d-1}=\frac{d}{r_0^{d+q}}\frac{N_L-1}{N-1}
\end{equation}
to write
\begin{equation}
\label{eq:I1}
I_1(N,d,q)
=dJ_1(d,q)\left(\frac{\tilde r_N}{r_0}\right)^{d+q}\frac{N_L-1}{N-1}
=\frac{q}{d+q}\left(\frac{\tilde r_N}{r_0}\right)^{d+q}\frac{N_L-1}{N-1}
\end{equation}
and
\begin{equation}
\label{eq:Il}
I_l(N,d,q)=I_1(N,d,q)\frac{J_l(d,q)}{J_1(d,q)}
\left(\frac{\mu_L}{(N-1)g\eta_{T}}\right)^{l-1}
\;.
\end{equation}

In the upper TF regime, i.e., $N\gg N_T$, we need to do integrals
over the TF probability density~(\ref{eq:upperthomasfermi}),
\begin{eqnarray}
\label{eq:A9}
K_l(N,d,q)&\equiv&
\int d^D\!\rho\,d^d\!r\,|\psi_N(\bm\rho,\bm r)|^{2l} \nonumber \\
& = & \left(\frac{k/2}{(N-1)g}\right)^l
\int d\Omega_{D-1} \int_0^{\tilde\rho_N} \rho^{D-1}d\rho\,
\int d\Omega_{d-1} \int_0^{\tilde r_N(\rho)} r^{d-1}dr\,(\tilde r_N^q(\rho) - r^q)^l \nonumber \\
& = & \left(\frac{m\omega_T^2/2}{(N-1)g}\right)^l
\left(\frac{m\omega_T^2}{k}\right)^{d/q}\tilde\rho_N^{D+2(l+d/q)}
S_{\!D-1}S_{\!d-1} J_{l+d/q}(D,2)J_l(d,q) \nonumber \\
& = & \left(\frac{\mu_N}{(N-1)g}\right)^l
\left(\frac{m\omega_T^2}{k}\right)^{d/q}\tilde\rho_N^{D+2d/q}
S_{\!D-1}S_{\!d-1} J_{l+d/q}(D,2)J_l(d,q)
\;.
\end{eqnarray}
Now we use $m\omega_T^2/k=r_0^{q+2}/4\rho_0^4$ and
\begin{equation}
\label{eq:A10}
\frac{m\omega_T^2/2}{(N-1)g}
\left(\frac{m\omega_T^2}{k}\right)^{d/q}=
\frac{1}{32\pi\beta_d2^{2d/q}}\frac{1}{\rho_0^{5-d+2d/q}}
\frac{N_T-1}{N-1}
\end{equation}
to write
\begin{equation}
\label{eq:K1}
K_1(N,d,q)=
dJ_{1+d/q}(D,2)J_1(d,q)
\frac{S_{\!D-1}}{4(4\pi)^{D/2}2^{2d/q}}
\left(\frac{\tilde\rho_N}{\rho_0}\right)^{5-d+2d/q}\frac{N_T-1}{N-1}
\end{equation}
and
\begin{equation}
K_l(N,d,q)=
K_1(N,d,q)\frac{J_{l+d/q}(d,q)J_l(d,q)}{J_{1+d/q}(d,q)J_1(d,q)}
\left(\frac{\mu}{(N-1)g}\right)^{l-1}\;.
\label{eq:Kl}
\end{equation}

\bibliography{paper}

\end{document}